\documentclass[11pt,a4paper]{article}

\usepackage{fix-cm}                       % Allow arbitrary font sizes
\usepackage{amsmath, amssymb, amsfonts}  % Standard math packages
\usepackage{graphicx}                     % For figures
\usepackage{hyperref}                     % For links
\usepackage{caption}                      % Captions for figures/tables
\usepackage{placeins}
\usepackage{microtype}
\usepackage{xurl}
\usepackage{bm}                           % Bold math symbols
  \usepackage[margin=1in]{geometry}  % Page layout

\usepackage[backend=biber, style=numeric, sorting=none]{biblatex}
\usepackage{float}

\numberwithin{equation}{section}

\begin{document}

  % Preprint format (article class)
  \title{Elliptic solutions, gauge normal forms and a degenerate four-wave-mixing projection for coherent couplers}
  \author{Graham Hesketh \\
  \small \texttt{gdh1e10@gmail.com}}
  \date{October 25, 2026}
  \maketitle

\begin{abstract}
We relate a complex Hamiltonian extension of the asymmetric conservative coherent coupler to three-mode degenerate four-wave mixing.
A trajectory-dependent reciprocal gauge preserves the modal products and gives an autonomous normal form with meromorphic envelopes; we construct a projection from degenerate four-wave mixing onto this normal form.
Building on established elliptic power-transfer and phase descriptions, we reconstruct the complex envelopes explicitly in Weierstrass $\sigma$ and $\zeta$ functions for unequal propagation constants and self-phase coefficients with symmetric cross-phase coupling.
This representation exposes the noninteger powers of sigma ratios introduced generically by unequal self-phase coefficients, and their removal by the gauge.
The lifted modes are exponentially gauged Kronecker kernels, equivalently genus-one Baker--Akhiezer functions, and the conserved Hamiltonian product relation is represented by a four-point Frobenius--Stickelberger determinant identity.
These mode-level relationships place the coupler within a wider family of integrable continuous-wave systems; numerical comparisons verify the formulae and their evaluation.
\end{abstract}

% ---------- SECTION -------------
% --------------------------------

\section{Introduction}

The nonlinear coherent coupler describes power exchange between two evanescently coupled waveguide modes with Kerr nonlinearity.
Jensen's analysis established its power-dependent switching behaviour and its role as a model for optical processing \cite{jensenNonlinearCoherentCoupler1982}.
Daino, Gregori and Wabnitz developed a Stokes-space description of Jensen's symmetric coupler, reduced its power imbalance to the Duffing equation, and analysed its equilibria, bifurcations and separatrices \cite{dainoStabilityAnalysisNonlinear1985}.
Their elliptic solution also determines the remaining Stokes components and hence the phase-invariant field products.
Peng and Ankiewicz derived the individual phase shifts for general two-input excitation of the symmetric Kerr coupler, using Jacobi elliptic functions and elliptic integrals \cite[Sections~2--5]{pengIntensitydependentPhaseShifts1990}.
Together with the powers, these phase solutions reconstruct the complex fields.

Analytical descriptions also exist for asymmetric couplers.
Snyder et al. treated unequal linear propagation constants and Kerr coefficients through power-flow portraits, conserved quantities and quadratures for the propagation \cite[Section~3.B and Appendix~A]{snyderPhysicsNonlinearFiber1991}.
Elliptic power-transfer solutions for single-core excitation were given for asymmetrically distributed Kerr nonlinearity by Karpierz \cite{karpierzNonlinearDirectionalCouplers1992}, and for linear mismatch with nonlinear asymmetry by Wang and Lee \cite{wangEffectsLinearMismatch2001}.
Related active systems have also been studied: Kominis, Bountis and Flach examined nonlinear supermodes and directed transport in an asymmetric coupler with gain and loss \cite{kominisAsymmetricActiveCoupler2016}.
Here we study an asymmetric conservative coupler and its complex Hamiltonian extension; we do not assume that this extension includes the general physical gain--loss system.

The established power-transfer and phase descriptions provide the analytical foundation for this work.
Our interest is in the complex analytic structure made visible by reconstructing the envelopes in Weierstrass notation and in transformations of the full mode equations.
The progression from $\wp$ to its primitive $-\zeta$, and then to sigma ratios through logarithmic derivatives, exposes the poles, zeros, exponential factors and possible branching of the modes.
It also permits a direct comparison with the functions used in other integrable systems.

The coupler belongs to a family of continuous-wave Hamiltonian mode systems with two interaction products and a polynomial in the modal products.
Related optical examples include quadratic wave mixing, polarisation dynamics and four-wave mixing \cite{heskethGeneralComplexEnvelope2015,heskethFWMWeierstrassElliptic2026}.
The interaction in the present two-mode coordinates is linear, whereas a lift to degenerate four-wave mixing has total interaction degree four, with one mode repeated.
This relationship concerns the full mode equations, beyond their scalar elliptic power reductions.
The related hyperelliptic construction \cite{zotero-item-855} extends the polynomial Hamiltonian setting to local sigma-function solutions on smooth odd-degree curves under stated regularity and degree conditions.
Understanding the genus-one case provides explicit examples of the function and determinant structures that occur in that less familiar setting.

\subsection{Our contribution}

Building on these established power-transfer and phase descriptions, we investigate the complex analytic structure of coherent-coupler envelopes and their relationship to degenerate four-wave mixing.
The central construction is a reciprocal gauge that preserves the modal products and gives an autonomous meromorphic normal form, together with a nonlinear projection from three-mode degenerate FWM onto that form.
The lifted modes have a common-pole sigma representation and can be identified explicitly with exponentially gauged Kronecker kernels, or genus-one Baker--Akhiezer functions.
This projection concerns the full mode equations and is distinct from a linear change between core and normal-mode bases.

To expose this structure, we give explicit complex-envelope formulae for unequal propagation and self-phase coefficients with symmetric cross-phase coefficients.
The formulae describe generic nonsingular elliptic trajectories with compatible initial data and locally chosen branches.
They identify the fractional-power factors introduced generically by unequal self-phase coefficients, and show how the reciprocal gauge removes them.
This branching under complex continuation is distinguished from the unique propagation of specified physical initial data along a regular real path.
We also represent the Hamiltonian product relation by the Frobenius--Stickelberger determinant, connecting the dynamics with classical elliptic addition identities.
These structural relationships are the principal motivation for the present work; the numerical comparisons verify the formulae without requiring a computational advantage over direct integration.

Sections~2--4 introduce the model, its physical realisation and its commuting conserved quantities.
Sections~\ref{sec:mode-power}--\ref{sec:mode-sols-u-v} reconstruct the powers and envelopes.
Section~\ref{sec:coord-transforms} gives the gauge and degenerate-FWM projection, and Section~\ref{app:fs-det-ham-proof} develops the determinant relation.
Numerical comparisons follow in Section~\ref{sec:num-val}, and Section~\ref{sec:discussion} discusses the Kronecker and Baker--Akhiezer descriptions.

% ---------- SECTION -------------
% --------------------------------

\section{Jensen's symmetric coherent coupler}
In this section we recall the coherent coupler system from Jensen's classic paper \cite{jensenNonlinearCoherentCoupler1982} that forms the starting point for our study. 
Jensen's system of ordinary differential equations, shown in \eqref{eq:jensen}, describes propagation in the quasi-continuous-wave limit, i.e., when temporal envelope variations are negligible compared to the spatial evolution along the propagation direction, allowing time derivatives to be neglected.
Physically, the equations govern the evolution of optical waves travelling in two waveguides placed close together, which exchange energy through coherent coupling of their evanescent fields.
This interaction is modelled by the linear coupling terms parameterised by $\Omega_2$.
Each wave also experiences self-phase modulation (SPM), where its own intensity induces a phase shift on itself, and cross-phase modulation (XPM), where the intensity of the other wave induces an additional phase shift. 
The system is described by the following coupled mode equations:

\begin{align}
\frac{dA_1}{dZ} &= i \left(|A_1|^2 {\Omega}_{3} + 2 |A_2|^2 {\Omega}_{4}\right) A_1 + i {\Omega}_{1} A_1  + i {\Omega}_{2} A_2, \notag \\
\frac{dA_2}{dZ} &= i \left(2 |A_1|^2 {\Omega}_{4} + |A_2|^2 {\Omega}_{3}\right) A_2 + i {\Omega}_{2} A_1  + i {\Omega}_{1} A_2, \label{eq:jensen}
\end{align}

where:
\begin{itemize}
    \item $A_j$ with $j=1,2$ are the slowly-varying complex field amplitudes
    \item $Z$ is the propagation distance
    \item $\Omega_1$ is the common propagation constant
    \item $\Omega_2$ is the linear coupling coefficient
    \item $\Omega_3$ is the self-phase modulation (SPM) coefficient
    \item $\Omega_4$ is the cross-phase modulation (XPM) coefficient.
\end{itemize}
We refer the reader to \cite{jensenNonlinearCoherentCoupler1982} for details on how to express the $\Omega_j$ coefficients in terms of material parameters and overlap integrals. 
Jensen states that the waveguides in \eqref{eq:jensen} were taken to be identical in order to simplify analysis and obtain solutions in standard elliptic integrals, hence the symmetry in the parameters in the two equations.
In this work, however, we analytically solve a generalised system that does not enforce this symmetry.
Jensen's system emerges as a special case.
We present the generalised system in the next section.

% ---------- SECTION -------------
% --------------------------------

\section{The generalised asymmetric coherent coupler}
In this section we normalise and generalise Jensen's system given in \eqref{eq:jensen} to arbitrary coefficients to broaden the scope of applicability.
We use the same notation as in \cite{heskethFWMWeierstrassElliptic2026} for a general FWM system, the only difference being that here there are two modes rather than four, so the mode index $j$ takes values $1,2$, whereas in the FWM case $j=1,2,3,4$. 
The system that we will consider is thus:
\begin{align}
\frac{d}{d z} u_j{\left(z \right)} &= - \left({a}_{j} + \sum_{k=1}^{2} {a}_{j,k}\,u_k v_k \right) u_j + \prod\limits_{k=1, k \ne j}^{2} v_k, \notag \\
\frac{d}{d z} v_j{\left(z \right)} &= \left({a}_{j} + \sum_{k=1}^{2} {a}_{j,k}\,u_k v_k \right) v_j - \prod\limits_{k=1, k \ne j}^{2} u_k, \label{eq:uv-system}
\end{align}
where we refer to the parameters $a_j, a_{j,k} \in \mathbb{C}$ as propagation constants and phase modulation parameters, respectively, and take $a_{2,1} = a_{1,2}$.
As there are only two modes, the product of modes governing wave mixing in \eqref{eq:uv-system} only contains one term and that term provides the linear coupling. 
Jensen's system in \eqref{eq:jensen} is recovered from \eqref{eq:uv-system} by introducing the original Jensen propagation coordinate $Z=z/\Omega_2$ and making the following substitutions:
\begin{align}
a_1 &= - a_2 = - \frac{i {\Omega}_{1}}{{\Omega}_{2}}, \notag \\
a_{1,1} &= a_{2,2} = \frac{i {\Omega}_{3}}{{\Omega}_{2}}, \notag \\
a_{1,2} &= a_{2,1} = - \frac{2 i {\Omega}_{4}}{{\Omega}_{2}}, \notag \\
u_1(z) &= A_1{\left(Z\right)} e^{- \frac{i \pi}{4}}, \notag \\
u_2(z) &= A^{*}_2{\left(Z\right)} e^{- \frac{i \pi}{4}}, \notag \\
v_1(z) &= -A^{*}_1{\left(Z\right)} e^{\frac{i \pi}{4}}, \notag \\
v_2(z) &= A_2{\left(Z\right)} e^{\frac{i \pi}{4}}, \label{eq:uv-to-jensen}
\end{align}
where $A^*$ denotes the complex conjugate and $z=\Omega_2 Z$ is the rescaled coordinate used in \eqref{eq:uv-system}.
Therefore, Jensen's system in \eqref{eq:jensen} is a special case of the more general system in \eqref{eq:uv-system}.

In the following sections we provide the complete analytic solution for the generalised system in \eqref{eq:uv-system}.
We treat $u_j,v_j$ as independent complex Hamiltonian variables, with physical fields obtained by imposing a preserved real slice.
For real Jensen coefficients and real $z$, \eqref{eq:uv-to-jensen} gives
\begin{equation}
v_1=-u_1^*,\qquad v_2=u_2^*,\qquad
u_1v_1=-|A_1|^2,\quad u_2v_2=|A_2|^2.
\label{eq:physical-slice}
\end{equation}
More generally, this slice is preserved when all $a_j,a_{j,k}$ are purely imaginary, allowing unequal propagation and self-phase coefficients.
The independent variables expose the complex Hamiltonian structure; on the physical slice they do not introduce extra independent optical fields.
In all that follows we refer to $u_j, v_j$ as modes, which we use as a general term for components of the coupled system. 
We refer to their product $u_j v_j$ as the modal power; while this quantity is not necessarily real-valued, the conceptual analogy is useful.

% This form encompasses the most general coupled system that the methods herein solve and may facilitate future extensions to related systems such as parity-time (PT) symmetric configurations such as the generalisation explored in \cite{heskethGeneralComplexEnvelope2015, sarmaContinuousDiscreteSchrodinger2014}.

% ---------- SECTION -------------
% --------------------------------

\section{Conserved quantities}

In this section we identify conserved quantities of the system. 
The approach and notation follow that of \cite{heskethFWMWeierstrassElliptic2026}, adjusted for two modes instead of four.
The system in \eqref{eq:uv-system} can be formulated as the following canonical Hamiltonian system:
\begin{align}
\label{eq:ham-uv}
H(u_1, u_2, v_1, v_2) &= -\sum_{j=1}^{2} a_{j}\, u_j v_j - \frac{1}{2}\,\sum_{j,k=1}^{2} a_{j,k}\, u_j v_j u_k v_k + \prod\limits_{j=1}^{2} u_j + \prod\limits_{j=1}^{2} v_j, \\
\frac{d}{d z} u_j{\left(z \right)} &= \frac{\partial H}{\partial v_j}, \\
\frac{d}{d z} v_j{\left(z \right)} &= -\frac{\partial H}{\partial u_j}, \\
\frac{d}{d z} H &= 0
\end{align}
The conservation of $H$ is to be expected as a consequence of \eqref{eq:ham-uv} lacking an explicit $z$ dependence. 
The pair $u_j, v_j$ are Hamiltonian conjugates, and each pair represents one of two degrees of freedom in the system. 
The modal power $u_j v_j$ evolves according to:
\begin{equation}
\label{eq:duv_j} 
\frac{d}{d z}\left(u_j v_j\right) = - \prod\limits_{k=1}^{2} u_k + \prod\limits_{k=1}^{2} v_k
\end{equation}

As the right-hand side of \eqref{eq:duv_j} is the same for all $j$, 
we may define constants $\gamma_j$ and a function $\rho(z)$ such that:
\begin{equation}
\label{eq:uvj-gam-rho}
u_j(z) v_j(z) = \gamma_j - \rho(z)
\end{equation}
from which it follows that there is an intermodal power conservation law of the form:
\begin{equation}
\label{eq:power-conserved}
u_1(z) v_1(z) - u_2(z) v_2(z) = \gamma_1 - \gamma_2.
\end{equation}

Writing $D=u_1v_1-u_2v_2$, the canonical Poisson bracket gives $\{H,D\}=0$.
The differentials $dH$ and $dD$ are independent on a generic open set: the flow generated by $D$ rescales the two $u$ modes oppositely and leaves both interaction products unchanged, while $H$ also depends on their sum.
Thus $H,D$ give two generically independent commuting integrals for this two-degree-of-freedom Hamiltonian system \cite{arnoldMathematicalMethodsClassical1989}.
On the physical slice \eqref{eq:physical-slice}, $D$ is minus the total optical power.
The constants $\gamma_j$ are determined only up to a common additive shift, since the transformation $\gamma_j\mapsto\gamma_j+c$ and $\rho(z)\mapsto\rho(z)+c$ leaves \eqref{eq:uvj-gam-rho} unchanged.
We choose to impose the constraint that:
\begin{equation}
\sum\limits_{j=1}^{2}\gamma_j = 0,
\end{equation} 
after which $\rho(z)$ is minus the mean modal power and $\gamma_j$ is the constant difference between a mode's modal power and the mean.

% ---------- SECTION -------------
% --------------------------------

\section{Solutions for modal powers in terms of Weierstrass \texorpdfstring{$\wp$}{wp} elliptic functions}\label{sec:mode-power}

In this section we give analytic solutions for the modal powers.
As in the previous section, the approach and notation closely follow that of \cite{heskethFWMWeierstrassElliptic2026}, adjusted for two modes instead of four.
The fundamental Weierstrass elliptic function theory used in this and subsequent sections can be found in \cite{whittakerCourseModernAnalysis2021}. 
In order to obtain elliptic function solutions in wave mixing dynamics, the key observation is that the derivative of the modal power in \eqref{eq:duv_j} is proportional to the difference of the two wave mixing product terms, and that the Hamiltonian in \eqref{eq:ham-uv} contains their sum.
We then note the simple but important identity:
\begin{equation}
\label{eq:prod-id}
\left(\prod\limits_{k=1}^{2} u_k - \prod\limits_{k=1}^{2} v_k\right)^2 - \left(\prod\limits_{k=1}^{2} u_k + \prod\limits_{k=1}^{2} v_k\right)^2 = - 4 \prod\limits_{j=1}^{2} u_j v_j
\end{equation}
which enables us to square \eqref{eq:duv_j} and replace wave mixing terms with phase modulation terms, i.e., monomials of $u_j v_j$ which can all be expressed in terms of $\rho(z)$ via \eqref{eq:uvj-gam-rho}.
To proceed in this manner, let us introduce the function $Q$ which represents the phase modulation part of the Hamiltonian and is thus itself expressible as a polynomial of $\rho(z)$:
\begin{align}
Q(u_1(z)v_1(z), u_2(z)v_2(z)) &= a_0 + \sum_{j=1}^{2} a_{j}\, u_j v_j + \frac{1}{2}\,\sum_{j,k=1}^{2} a_{j,k}\, u_j v_j u_k v_k , \notag  \\
& = \sum_{l=0}^{2}b_l\,\rho\left(z\right)^l \label{eq:Q-uv}
\end{align}
where $a_0=H$ is introduced so that the conserved Hamiltonian appears as the constant term of the polynomial. Observe that squaring \eqref{eq:duv_j} and substituting \eqref{eq:ham-uv}, \eqref{eq:uvj-gam-rho}, \eqref{eq:prod-id}, and \eqref{eq:Q-uv} gives:
\begin{align}
\left(\frac{d}{d z} \rho{\left(z \right)}\right)^{2} &= Q^{2}{\left({\gamma}_{1} - \rho{\left(z \right)}, {\gamma}_{2}  - \rho{\left(z \right)}\right)} - 4 \prod_{j=1}^{2} \left({\gamma}_{j} - \rho{\left(z \right)}\right), \label{eq:drho-sqrd-1} \\
\left(\frac{d}{d z} \rho{\left(z \right)}\right)^{2} &=  \left(\sum_{l=0}^{2}b_l\,\rho\left(z\right)^l\right)^{2} - 4 \prod_{j=1}^{2} \left({\gamma}_{j} - \rho{\left(z \right)}\right), \label{eq:drho-sqrd-2} \\
\left(\frac{d}{d z} \rho{\left(z \right)}\right)^{2} &= \sum_{l=0}^{4}d_l\,\rho\left(z\right)^l, \label{eq:drho-sqrd-3} \\
\left(\frac{d}{d z} \rho{\left(z \right)}\right)^{2} &= d_4 \prod_{l=1}^{4} \left(\rho\left(z\right) -  \lambda_l\right), \label{eq:drho-sqrd-4}
\end{align}
where $b_l$ and $d_l$ are given in terms of other parameters and initial conditions in Appendices \ref{app:param-def-appendix} and \ref{app:init-conds}, and where $\lambda_l$ are the roots of the quartic polynomial in $\rho(z)$, i.e.:
\begin{equation}
\sum_{k=0}^{4}d_k\,\lambda_{l}^k = 0 \label{eq:lambda-root}
\end{equation}
Equation \eqref{eq:drho-sqrd-4} has elliptic solutions when the quartic is nonsingular.
The displayed root-based chart assumes $b_2\ne0$, a simple chosen root $\lambda_1$, and nonzero denominators at the evaluation points.
If $b_2=0$, the scalar equation has degree at most two and is treated by elementary integration; repeated-root cases require separate limits.
Zeros and poles of individual chart expressions can require a different local representation. 
By separating phase modulation and wave mixing terms in the Hamiltonian, and by exploiting intermodal power conservation, we obtain the scalar quartic equation directly from the conservation laws. Its integration is the classical elliptic reduction; fixing its initial-data points still requires inversion of $\wp$, as described below.
To solve it, we now transform \eqref{eq:drho-sqrd-4} from quartic to the standard cubic form of the Weierstrass $\wp$ function using the classical trick which can be conceptualised in three steps \cite{whittakerCourseModernAnalysis2021}:
\begin{equation}
\label{eq:wp-4-to-3}
\begin{array}{l rcl}
\text{(1) shift by any root so RHS is 0 when $q(z)=0$} 
& \rho(z) &=& q(z) + \lambda_1,\\
\text{(2) invert to reduce the quartic to a cubic} 
& s(z) &=& \dfrac{1}{q(z)},\\
\text{(3) shift and scale to match Weierstrass coefficients} 
& w(z) &=& C_1 s(z) + C_0.
\end{array}
\end{equation}

For definiteness and without loss of generality, we choose the root $\lambda_1$. 
The procedure sketched in \eqref{eq:wp-4-to-3} is then implemented in the following single transformation:
\begin{align}
\rho{\left(z \right)} &= {\lambda}_{1} + \frac{d_4}{- 4 w{\left(z \right)} \prod_{l=1}^{3} {\Delta}_{l}  + \frac{d_4}{3} \sum_{l=1}^{3} {\Delta}_{l} }, \label{eq:rho-wp-1} \\
\Delta_l &= \frac{1}{\lambda_{l+1} - \lambda_1}, \notag \\
\left(\frac{d}{d z} w{\left(z \right)}\right)^{2} &= 4\,w{\left(z \right)}^3 - g_2\,w{\left(z \right)} - g_3, \label{eq:dwp} \\
g_{2} &= {d}_{0} {d}_{4} - \frac{{d}_{1} {d}_{3}}{4} + \frac{{d}_{2}^{2}}{12}, \label{eq:g2-d} \\
g_{3} &= \frac{{d}_{0} {d}_{2} {d}_{4}}{6} - \frac{{d}_{0} {d}_{3}^{2}}{16} - \frac{{d}_{1}^{2} {d}_{4}}{16} + \frac{{d}_{1} {d}_{2} {d}_{3}}{48} - \frac{{d}_{2}^{3}}{216} \label{eq:3-d}
\end{align}
where the constants $g_2$ and $g_3$ are known as Weierstrass elliptic invariants. 
Equation \eqref{eq:dwp} defines the Weierstrass elliptic $\wp$ function and the solution is:
\begin{equation}
w(z) = \wp (z - z_0, g_2, g_3)
\end{equation}
where $z_0$ is a constant chosen to match initial conditions. 
The constant $z_0$, and the other special points introduced below, can be obtained by inverting $\wp$ using an elliptic integral \cite{whittakerCourseModernAnalysis2021}. 
As $\wp$ is an even function, it is necessary to also specify a corresponding condition for the derivative when inverting to fix the sign, i.e., to find $z$ from known $x,y$ we give conditions such as $\wp\left(z\right)=x, \wp'\left(z\right)=y$, where $\wp'$ is the derivative of $\wp$. 
From elliptic function theory, the points obtained during such an inversion are determined modulo the period lattice of the doubly periodic $\wp$ function.
In addition to $z_0$, we introduce two other types of special points that characterize the solution structure and enable the use of elliptic function identities to solve for the modes in Section \ref{sec:mode-sols-u-v}.
The point $z_0+z_1$ corresponds to the pole of $\rho(z)$ where the denominator in \eqref{eq:rho-wp-1} vanishes.
The points $\mu_j$ are the locations where individual modal powers vanish, $u_j(\mu_j)v_j(\mu_j) = 0$, and thus where $\rho(\mu_j) = \gamma_j$ from \eqref{eq:uvj-gam-rho}.
These special points are defined implicitly through their Weierstrass function values:
\begin{align}
\wp{\left(z_{0}\right)} &= \frac{{d}_{2}}{12} + \frac{{d}_{3} {\lambda}_{1}}{4} + \frac{{d}_{4} {\lambda}_{1}^{2}}{2} + \frac{- {d}_{1} - 2 {d}_{2} {\lambda}_{1} - 3 {d}_{3} {\lambda}_{1}^{2} - 4 {d}_{4} {\lambda}_{1}^{3}}{4 \left(- \rho{\left(0 \right)} + {\lambda}_{1}\right)}, \notag \\
\wp'{\left(z_{0}\right)} &= \frac{\left({d}_{1} + 2 {d}_{2} {\lambda}_{1} + 3 {d}_{3} {\lambda}_{1}^{2} + 4 {d}_{4} {\lambda}_{1}^{3}\right) }{4 \left(\rho{\left(0 \right)} - {\lambda}_{1}\right)^{2}}\left. \frac{d}{d z} \rho{\left(z \right)} \right|_{\substack{ z=0 }}, \notag \\
\wp{\left(z_{1}\right)} &= \frac{{d}_{2}}{12} + \frac{{d}_{3} {\lambda}_{1}}{4} + \frac{{d}_{4} {\lambda}_{1}^{2}}{2}, \notag \\
\wp'{\left(z_{1}\right)} &= \frac{\left(- {d}_{1} - 2 {d}_{2} {\lambda}_{1} - 3 {d}_{3} {\lambda}_{1}^{2} - 4 {d}_{4} {\lambda}_{1}^{3}\right) b_2}{4}, \notag \\
\wp{\left({\mu}_{j} - z_{0}\right)} &= \frac{{d}_{2}}{12} + \frac{{d}_{3} {\lambda}_{1}}{4} + \frac{{d}_{4} {\lambda}_{1}^{2}}{2} - \frac{- {d}_{1} - 2 {d}_{2} {\lambda}_{1} - 3 {d}_{3} {\lambda}_{1}^{2} - 4 {d}_{4} {\lambda}_{1}^{3}}{4 \left({\gamma}_{j} - {\lambda}_{1}\right)}, \notag \\
\wp'{\left({\mu}_{j}- z_{0}\right)} &= - \frac{\left({b}_{0} + {b}_{1} {\gamma}_{j} + {b}_{2} {\gamma}_{j}^{2}\right) \left({d}_{1} + 2 {d}_{2} {\lambda}_{1} + 3 {d}_{3} {\lambda}_{1}^{2} + 4 {d}_{4} {\lambda}_{1}^{3}\right)}{4 \left({\gamma}_{j} - {\lambda}_{1}\right)^{2}}. \label{eq:wp-points}
\end{align}
The solutions for modal powers are then found from \eqref{eq:uvj-gam-rho} and \eqref{eq:rho-wp-1} to be:
\begin{align}
u_j{\left(z\right)} v_j{\left(z\right)} &= \rho(\mu_j) - \rho(z), \notag \\
&= \frac{\wp'{\left(z_{1}\right)}}{ b_2 \left(\wp{\left(z_{1}\right)} - \wp{\left({\mu}_{j} - z_{0}\right)}\right)} - \frac{\wp'{\left(z_{1}\right)}}{b_2 \left(\wp{\left(z_{1}\right)} - \wp{\left(z - z_{0}\right)}\right) }, \notag \\
&= \frac{\wp'{\left(z_{1}\right)} }{ b_2 \left(\wp{\left(z_{1}\right)} - \wp{\left({\mu}_{j}- z_{0}\right)}\right)} \frac{ \left(\wp{\left(z - z_{0}\right)} - \wp{\left({\mu}_{j} - z_{0}\right)}\right) }{ \left(\wp{\left(z - z_{0}\right)} - \wp{\left(z_{1}\right)}\right) }. \label{eq:uv-wp}
\end{align}

% ---------- SECTION -------------
% --------------------------------

\section{Solutions for modes in terms of Weierstrass \texorpdfstring{$\sigma$}{sigma}, \texorpdfstring{$\zeta$}{zeta} functions} \label{sec:mode-sols-u-v}

Having obtained the modal power solutions in the previous section, we now proceed to find solutions for the individual modes $u_j(z)$ and $v_j(z)$ themselves.
Direct substitution of \eqref{eq:ham-uv}, \eqref{eq:duv_j}, \eqref{eq:uvj-gam-rho}, and \eqref{eq:power-conserved} into \eqref{eq:uv-system} shows that the equations can be recast in terms of logarithmic derivatives:
\begin{align}
\frac{\frac{\partial}{\partial z} u_j{\left(z\right)}}{u_j{\left(z\right)}} &= \frac{1}{2}\frac{\rho'{\left(z \right)} - \rho'{\left({\mu}_{j} \right)}}{\rho{\left(z \right)} - \rho{\left({\mu}_{j} \right)}} + \rho{\left(z \right)} {\Lambda}_{1,j} + {\Lambda}_{0,j}, \notag \\
\frac{\frac{\partial}{\partial z} v_j{\left(z\right)}}{v_j{\left(z\right)}} &= \frac{1}{2}\frac{\rho'{\left(z \right)} + \rho'{\left({\mu}_{j} \right)}}{\rho{\left(z \right)} - \rho{\left({\mu}_{j} \right)}} - \rho{\left(z \right)} {\Lambda}_{1,j} - {\Lambda}_{0,j}, \label{eq:dlog-u-v-rho} \\
{\Lambda}_{0,j} &= - {a}_{j} - \frac{{\gamma}_{j}}{4}\sum_{\substack{1 \leq k \leq 2\\1 \leq l \leq 2}} {a}_{k,l} - \sum_{l=1}^{2} {a}_{j,l} {\gamma}_{l} + \frac{1}{2}\sum_{k=1}^{2} {a}_{k,k} {\gamma}_{k} + \sum_{k=1}^{2} \frac{{a}_{k}}{2}, \\
{\Lambda}_{1,j} &= \sum_{k=1}^{2} {a}_{j,k} - \frac{1}{4}\sum_{\substack{1 \leq k \leq 2\\1 \leq l \leq 2}}{a}_{k,l},
\end{align}
where the right-hand side is expressed purely in terms of $\rho$ and its derivative $\rho'$.
This logarithmic derivative form is advantageous because it can be integrated directly to yield solutions in terms of Weierstrass $\sigma$ functions, as we now show.

We substitute the modal power solution \eqref{eq:uv-wp} into \eqref{eq:dlog-u-v-rho} and apply the elliptic function identity:
\begin{equation}
\frac{\wp'{\left(x,g_{2},g_{3} \right)}}{\wp{\left(x,g_{2},g_{3} \right)} - \wp{\left(y,g_{2},g_{3} \right)}} = \zeta{\left(x + y,g_{2},g_{3} \right)} + \zeta{\left(x - y,g_{2},g_{3} \right)} - 2 \zeta{\left(x,g_{2},g_{3} \right)}
\end{equation}
which allows us to express the right-hand side in terms of Weierstrass $\zeta$ functions and constants:
\begin{align}
\frac{\frac{\partial}{\partial z} u_j{\left(z \right)}}{u_j{\left(z \right)}} =& \frac{\left(\zeta{\left(z - z_{0} + z_{1} \right)} - 2 \zeta{\left(z_{1} \right)} - \zeta{\left(z - z_{0} - z_{1} \right)}\right) {\Lambda}_{1,j}}{b_2} \nonumber \\[6pt]
& + \zeta{\left(z - 2 z_{0} + {\mu}_{j} \right)} - \frac{\zeta{\left(z - z_{0} - z_{1} \right)}}{2} - \frac{\zeta{\left(z - z_{0} + z_{1} \right)}}{2} \nonumber \\[6pt]
& - \frac{\zeta{\left({\mu}_{j} - z_{0} - z_{1}\right)}}{2} - \frac{\zeta{\left({\mu}_{j} - z_{0} + z_{1}\right)}}{2} + {\Lambda}_{0,j} + {\Lambda}_{1,j} {\lambda}_{1} \notag \\[12pt]
\frac{\frac{\partial}{\partial z} v_j{\left(z \right)}}{v_j{\left(z \right)}} =& - \frac{\left(\zeta{\left(z - z_{0} + z_{1} \right)} - 2 \zeta{\left(z_{1} \right)} - \zeta{\left(z - z_{0} - z_{1} \right)}\right) {\Lambda}_{1,j}}{b_2} \nonumber \\[6pt]
& + \zeta{\left(z - {\mu}_{j} \right)} - \frac{\zeta{\left(z - z_{0} - z_{1} \right)}}{2} - \frac{\zeta{\left(z - z_{0} + z_{1} \right)}}{2} \nonumber \\[6pt]
& + \frac{\zeta{\left({\mu}_{j} - z_{0} - z_{1}\right)}}{2} + \frac{\zeta{\left({\mu}_{j} - z_{0} + z_{1} \right)}}{2} - {\Lambda}_{0,j} - {\Lambda}_{1,j} {\lambda}_{1}. \label{eq:dlog-u-v-zeta}
\end{align}
Equations \eqref{eq:dlog-u-v-zeta} can be integrated by noting that the Weierstrass $\zeta$ function is the logarithmic derivative of the Weierstrass $\sigma$ function:
\begin{equation}
\label{eq:zeta-dlog_sigma}
\zeta\left(z, g_2, g_3\right)=\frac{\partial}{\partial z} \log{\left(\sigma\left(z, g_2, g_3\right)\right)}.
\end{equation}
Performing the integration and taking exponentials gives an explicit representation of the complex mode envelopes $u_j,v_j$ that will be used to analyse their branching and normal form:
\begin{align}
u_j{\left(z\right)} &= \frac{\alpha_j\,\sqrt{W_j} \,\sigma{\left(z - 2 z_{0} + {\mu}_{j} \right)} \exp\left(z {r}_{0,j} + \log{\left(\frac{\sigma{\left(z - z_{0} + z_{1} \right)}}{\sigma{\left(z - z_{0} - z_{1} \right)}} \right)} {r}_{1,j}\right)}{\sqrt{\wp{\left(z_{1} \right)} - \wp{\left(z - z_{0} \right)}} \sigma{\left({\mu}_{j} - z_{0}\right)} \sigma{\left(z - z_{0} \right)}}, \notag \\[12pt]
v_j{\left(z\right)} &= \frac{\sqrt{W_j} \,\sigma{\left(z - {\mu}_{j} \right)} \exp\left(- z {r}_{0,j} - \log{\left(\frac{\sigma{\left(z - z_{0} + z_{1} \right)}}{\sigma{\left(z - z_{0} - z_{1} \right)}} \right)} {r}_{1,j} \right)}{\alpha_j\,\sqrt{\wp{\left(z_{1} \right)} - \wp{\left(z - z_{0} \right)}} \sigma{\left({\mu}_{j} - z_{0}\right)} \sigma{\left(z - z_{0}\right)}} \label{eq:u-v-quartic}
\end{align}
or equivalently:
\begin{align}
u_j{\left(z\right)} &= \frac{\kappa\, \alpha_j\,\sqrt{W_j} \,\sigma{\left(z_1\right)}\sigma{\left(z - 2 z_{0} + {\mu}_{j} \right)} \exp\left(z {r}_{0,j} + \log{\left(\frac{\sigma{\left(z - z_{0} + z_{1} \right)}}{\sigma{\left(z - z_{0} - z_{1} \right)}} \right)} \left({r}_{1,j} - \frac{1}{2}\right)\right)}{ \sigma{\left({\mu}_{j} - z_{0}\right)} \sigma{\left(z - z_{0}  - z_1\right)}}, \notag \\[12pt]
v_j{\left(z\right)} &= \frac{\sqrt{W_j} \,  \sigma{\left(z_1\right)}\sigma{\left(z - {\mu}_{j} \right)} \exp\left(- z {r}_{0,j} - \log{\left(\frac{\sigma{\left(z - z_{0} + z_{1} \right)}}{\sigma{\left(z - z_{0} - z_{1} \right)}} \right)}  \left({r}_{1,j} - \frac{1}{2}\right) \right)}{\kappa \,\alpha_j \sigma{\left({\mu}_{j} - z_{0}\right)} \sigma{\left(z - z_{0} + z_1\right)}} \label{eq:u-v-quartic-no-wp-sqrt}
\end{align}
where the branch of the logarithm is chosen continuously along the integration path from $z=0$, $\alpha_j$ is the integration constant that can be fixed by initial conditions to capture any phase offset between a mode and its Hamiltonian conjugate, and square-root branches are chosen consistently with the same initial conditions.
Only the relative normalisation constants are free once the modal products, $H$ and $\rho'(0)$ have been specified; the chosen initial fields fix them and their product compatibility.
The other constants are:
\begin{align}
W_j &= \frac{\wp'{\left(z_{1} \right)}}{b_2\,\left(\wp{\left(z_{1} \right)} - \wp{\left({\mu}_{j} - z_{0}\right)}\right)}, \notag \\[6pt]
r_{0,j} &= {\Lambda}_{0,j} + {\Lambda}_{1,j} {\lambda}_{1} - \frac{2 \zeta{\left(z_{1} \right)} {\Lambda}_{1,j}}{b_2} - \frac{\zeta{\left({\mu}_{j} - z_{0} - z_{1}\right)}}{2} - \frac{\zeta{\left({\mu}_{j} - z_{0} + z_{1}\right)}}{2}, \notag \\[6pt]
r_{1,j} &= \frac{{\Lambda}_{1,j}}{b_2}, \notag \\[6pt]
\kappa &= \frac{\sqrt{\wp{\left(z_{1}\right)} - \wp{\left(z_{0}\right)}} \sigma{\left(z_{0}\right)} \sigma{\left(z_{1}\right)}}{\sqrt{\frac{\sigma{\left(z_{0} - z_{1}\right)}}{\sigma{\left(z_{0} + z_{1}\right)}}} \sigma{\left(z_{0} + z_{1}\right)}} = \pm 1
\end{align}
and it can be shown that:
\begin{equation}
  \sum_{j=1}^{2}r_{1,j} = 1. \label{eq:r1j-sum}
\end{equation}
These formulae apply locally on the nonsingular chart specified above.
Continuous continuation of the logarithm, rather than independent principal values at each point, is required along a propagation path.
For $b_2\ne0$, their fractional-power exponents satisfy
\begin{equation}
r_{1,1}-\frac12=\frac{a_{1,1}-a_{2,2}}{a_{1,1}+2a_{1,2}+a_{2,2}},\qquad
r_{1,2}-\frac12=-\frac{a_{1,1}-a_{2,2}}{a_{1,1}+2a_{1,2}+a_{2,2}}.
\label{eq:branch-asymmetry}
\end{equation}
Thus unequal self-phase coefficients generically produce branching under complex continuation, whereas unequal propagation constants alone do not.
Special integer exponents also give meromorphic envelopes.

We show in Section~\ref{app:fs-det-ham-proof} that the Hamiltonian \eqref{eq:ham-uv} can be expressed as the Frobenius--Stickelberger determinant, a classical identity relating products of Weierstrass $\sigma$ functions to determinants of derivatives of $\wp$.
This represents the conserved Hamiltonian relation in terms of classical elliptic addition identities.

\subsection{Jensen's special case}

In Jensen's case, there is notable simplification in the solutions as the waveguides are made identical.
In our notation, this translates into additional symmetries among $a_j$ and $a_{j,k}$ parameters as specified by their relation to Jensen's parameters in \eqref{eq:uv-to-jensen}.
These symmetries further translate into $b_1=0$, $2\Lambda_{1,j}=b_2$, and importantly $r_{1,j} = \frac{1}{2}$, which in \eqref{eq:u-v-quartic-no-wp-sqrt} can be seen to eliminate the logarithmic term in the exponentials, thereby removing the branching behaviour, reducing the solutions to a ratio of sigma functions, and thus, meromorphic functions.
In Section~\ref{sec:coord-transforms} we show that the branching behaviour can also be removed in the general case by a reciprocal gauge transformation.

% ---------- SECTION -------------
% --------------------------------

\section{Relating the coherent coupler to degenerate FWM} \label{sec:coord-transforms}

In this section we construct the mode-level relationship central to this paper: a reciprocal gauge from the coherent coupler to an autonomous meromorphic normal form, and a nonlinear projection from degenerate FWM onto that form.
The structure of the analytic solutions in \eqref{eq:u-v-quartic-no-wp-sqrt} naturally suggests this transformation.
We proceed in two steps: first we remove cross-phase modulation by a gauge transformation, and then we identify the resulting two-mode system as a projection of a degenerate FWM system.

\subsection{Removing cross-phase modulation with a gauge transform}

To begin, let us consider a transform of the system in \eqref{eq:uv-system} of the following form:
\begin{align}
u_j(z) &= \hat{u}_j(z) e^{- \phi_j(z)}, \notag \\
v_j(z) &= \hat{v}_j(z) e^{\phi_j(z)}, \\ 
\sum_{j=1}^{2} \phi_j(z) &= 0,
\end{align}
where transforming the Hamiltonian conjugate with the opposing phase leaves modal powers unchanged, and where the zero-sum condition on $\phi_j$ ensures that no additional exponential factors appear in the wave-mixing products.
We make the following choice for $\phi_j$ composed of a part linear in $z$ (labelled $L$) and a part nonlinear in $z$ (labelled $NL$) (where $L$ and $NL$ are function labels not exponents):
\begin{align}
N &= 2, \\
\phi_j(z) &= \phi_j^{L}(z) + \phi_j^{NL}(z) \\
\phi_j^{L}(z) &= z\left({a}_{j} - \frac{1}{N}\sum_{k=1}^{N} a_{k} - \frac{{\gamma}_{j}}{N} \sum\limits_{l,k=1}^{N} a_{l,k}\right), \\
\phi_j^{NL}(z)  &= \sum_{k=1}^{N} \left({a}_{j,k} - \frac{1}{N}\sum_{l=1}^{N} {a}_{l,k}\right) \int_0^z \hat{u}_k (\xi) \hat{v}_k (\xi) \, d\xi, \\
\sum_{j=1}^{N} \phi_j(z) &= \sum_{j=1}^{N} \phi_j^{L}(z) = \sum_{j=1}^{N} \phi_j^{NL}(z) =0.
\end{align}

This transforms \eqref{eq:uv-system} into the following system:
\begin{align}
\frac{\partial}{\partial z} \hat{u}_j &= \left(b_1 - 2 b_2\hat{u}_j \hat{v}_j \right) \frac{\hat{u}_j}{N} + \prod\limits_{k=1, k \ne j}^{N} \hat{v}_k \notag \\
\frac{\partial}{\partial z} \hat{v}_j &= \left( - b_1 + 2 b_2\hat{u}_j \hat{v}_j\right) \frac{\hat{v}_j}{N} - \prod\limits_{k=1, k \ne j}^{N} \hat{u}_k \label{eq:uv-hat-system}
\end{align}
for which the conserved canonical Hamiltonian is:
\begin{align}
\hat{H} &= \prod_{l=1}^{N} \hat{u}_l + \prod_{l=1}^{N} \hat{v}_l - \frac{1}{N}\sum_{l=1}^{N} \left(b_2 \hat{u}^{2}_l \hat{v}^{2}_l  - b_1 \hat{u}_l \hat{v}_l \right)
\end{align}
and the intermodal power conservation law is unchanged such that $\hat{u}_1 \hat{v}_1 - \hat{u}_2 \hat{v}_2 = \gamma_1 - \gamma_2$. 
To maintain the connection with the original coordinate system the value of $\hat{H}$ would be fixed at:
\begin{equation}
\hat{H} =  - \frac{\left({\gamma}_{1} - {\gamma}_{2}\right)^{2} {b}_{2}}{4} + {b}_{0}.
\end{equation}
Comparing \eqref{eq:uv-hat-system} with the original system \eqref{eq:uv-system}, we observe significant simplification.
All modes now share a single propagation constant $b_1/N$ and a single self-phase modulation coefficient $2b_2/N$, while all cross-phase modulation terms $a_{j,k}$ with $j \ne k$ have been eliminated.
The parameters $b_1$ and $b_2$ depend on the conserved constants $\gamma_j$; in the transformed coordinates they therefore encode the chosen power level as well as fibre parameters.
In terms of the solutions to the system, the effect of the $\phi_j^{NL}$ transform is to remove the $z$-dependent log terms weighted by $r_{1,j}$ in \eqref{eq:u-v-quartic}. 
The gauge integrates accumulated phase-modulation terms along each trajectory and applies reciprocal factors to the paired modes.
It is a change of representation for the autonomous equations on a fixed conserved-power leaf; we do not assert that it is a canonical coordinate transformation on the whole phase space.
On the physical slice its factors are phases and preserve the optical powers.
The solutions are:
\begin{align}
\hat{u}_j{\left(z\right)} &= \frac{{\hat{\alpha}}_{j} \sqrt{W_j}\sigma{\left(z_{1}\right)} \sigma{\left(z - 2 z_{0} + {\mu}_{j}\right)} e^{-z\left(\zeta\left(\mu_j + z_1 - z_0\right) + b_2\gamma_j\right)} }{ \sigma{\left({\mu}_{j} - z_{0}\right)} \sigma{\left(z - z_{0} - z_{1}\right)}}, \notag \\
\hat{v}_j{\left(z\right)} &= \frac{\sqrt{W_j}\sigma{\left(z_{1}\right)} \sigma{\left(z - {\mu}_{j}\right)} e^{z\left(\zeta\left(\mu_j + z_1 - z_0\right) + b_2\gamma_j\right)}}{ {\hat{\alpha}}_{j} \sigma{\left({\mu}_{j} - z_{0}\right)} \sigma{\left(z - z_{0} + z_{1}\right)}}, \label{eq:uhat-vhat-gauge-sols}
\end{align}
where $\hat{\alpha}_{j}$ are integration constants used to fix initial conditions, the elliptic functions use invariants $g_2, g_3$, and $\mu_1, \mu_2, z_0, z_1, g_2, g_3$ share the same definitions and values as those given in Section~\ref{sec:mode-power} for the coherent coupler, thereby connecting the two systems. 
The hatted modes $\hat{u}_j$ have poles at $z=z_0+z_1$ modulo the lattice, whereas their Hamiltonian conjugates $\hat{v}_j$ have poles at $z=z_0-z_1$. 
In other work on FWM \cite{heskethFWMWeierstrassElliptic2026}, we encountered cases where modes and their conjugates share poles at $z=z_0$ and there the modal power is the difference of two Weierstrass $\wp$ functions. 
This suggests rewriting the ratios in \eqref{eq:uhat-vhat-gauge-sols} as quotients of functions with a common pole at $z=z_0$:
\begin{align}
  \frac{ \sigma{\left(z - 2 z_{0} + {\mu}_{j}\right)} }{\sigma{\left(z - z_{0} - z_{1}\right)}} &= \frac{\left(\frac{\sigma{\left(z - 2 z_{0} + {\mu}_{j}\right)}}{\sigma{\left(z - z_{0}\right)}}\right)}{\left(\frac{\sigma{\left(z - z_{0} - z_{1}\right)}}{\sigma{\left(z - z_{0}\right)}}\right)}, \notag \\
  \frac{ \sigma{\left(z - {\mu}_{j}\right)} }{\sigma{\left(z - z_{0} + z_{1}\right)}} &= \frac{\left(\frac{\sigma{\left(z - {\mu}_{j}\right)}}{\sigma{\left(z - z_{0}\right)}}\right)}{\left(\frac{\sigma{\left(z - z_{0} + z_{1}\right)}}{\sigma{\left(z - z_{0}\right)}}\right)}, \label{eq:two-to-three-ratios-sig}
\end{align}
The sigma-ratio factors on the left of \eqref{eq:two-to-three-ratios-sig} are those appearing in the two hatted modes and their conjugates in \eqref{eq:uhat-vhat-gauge-sols}, while the factors in parentheses on the right can be interpreted as three modes and their Hamiltonian conjugates, with the denominator factors conjugate to one another.
The associated three-mode system is the degenerate FWM system described next.

\subsection{Projecting from degenerate FWM to the coherent coupler}\label{subsec:deg-fwm}
We now show how the coherent coupler system in \eqref{eq:uv-system} arises as a projection of a degenerate four-wave mixing (FWM) system from nonlinear fibre optics.

To start, let us recall that in \cite{heskethFWMWeierstrassElliptic2026} we showed how quasi-continuous-wave FWM systems can be transformed into the following parameterless non-degenerate Hamiltonian normal form of four modes $\tilde{u}_j(\xi), j=1,2,3,4$ with $\xi$ a propagation length variable, and their Hamiltonian conjugates $\tilde{v}_j(\xi)$ (note the tilde notation):
\begin{align}
\tilde{H} &= \sum_{l=1}^{4} \left(\frac{1}{8} \tilde{u}^{2}_l \tilde{v}^{2}_l  - \tilde{u}_l \tilde{v}_l \right) - \frac{1}{4}\left(\prod_{l=1}^{4} \tilde{u}_l + \prod_{l=1}^{4} \tilde{v}_l\right), \label{eq:H-deg-tilde}
\end{align}
\begin{align}
\frac{\partial}{\partial \xi} \tilde{u}_j &= \frac{\partial}{\partial \tilde{v}_j} \tilde{H} = -\left(1 - \frac{\tilde{u}_j \tilde{v}_j}{4} \right) \tilde{u}_j - \frac{1}{4}\prod\limits_{k=1, k \ne j}^{4} \tilde{v}_k, \notag \\
\frac{\partial}{\partial \xi} \tilde{v}_j &= -\frac{\partial}{\partial \tilde{u}_j} \tilde{H} = \left( 1 - \frac{\tilde{u}_j \tilde{v}_j}{4}\right) \tilde{v}_j + \frac{1}{4}\prod\limits_{k=1, k \ne j}^{4} \tilde{u}_k, \label{eq:uv-tilde-deg-system}
\end{align}
which conserves the Hamiltonian $\tilde{H}$ and, for constants  $\tilde{\gamma}_j$, three independent intermodal power differences: fixing $k=1$ and taking $j=2,3,4$ gives a basis, while all six unordered pairwise differences follow from it:
\begin{equation}
\label{eq:deg-fwm-power-conserved}
\tilde{u}_j(\xi) \tilde{v}_j(\xi) - \tilde{u}_k(\xi) \tilde{v}_k(\xi) = \tilde{\gamma}_j - \tilde{\gamma}_k.
\end{equation}

We now consider a degeneration and rescaling of this FWM system defined by the following substitutions in the Hamiltonian in \eqref{eq:H-deg-tilde} (note the bar notation on the right and the tilde notation on the left):
\begin{align}
  \chi\xi &= z, \notag \\
  \tilde{u}_1(\xi) &= \varsigma \sqrt{2}\chi \,\bar{u}_1(z), \notag \\
  \tilde{u}_2(\xi) &= \sqrt{2} \chi  \,\bar{u}_2(z), \notag \\
  \tilde{u}_3(\xi) &= \chi  \,\bar{u}_3(z), \notag \\
  \tilde{u}_4(\xi) &= \chi  \,\bar{u}_3(z), \notag \\
  \tilde{v}_1(\xi) &= \varsigma \sqrt{2} \chi  \,\bar{v}_1(z), \notag \\
  \tilde{v}_2(\xi) &= \sqrt{2} \chi  \,\bar{v}_2(z), \notag \\
  \tilde{v}_3(\xi) &= \chi  \,\bar{v}_3(z), \notag \\
  \tilde{v}_4(\xi) &= \chi  \,\bar{v}_3(z), \notag \\
\end{align}
where $\varsigma = \pm 1$, $\chi$ is a free parameter, and $\tilde{u}_3$ and $\tilde{u}_4$ coalesce to $\bar{u}_3$ (likewise for the corresponding conjugates) to create degeneracy and reduce four degrees of freedom (modes) to three.
We then divide the Hamiltonian by $2 \chi^3$.
The division by $2\chi^3$ has two parts. The factor $2\chi^2$ comes from the rescaling of the dependent variables as differentiating the substituted Hamiltonian with respect to $\bar u_j$ or $\bar v_j$ brings down the same factor that appears in the product of the corresponding $\tilde u_j,\tilde v_j$ scalings. 
Dividing by this factor restores the canonical Hamilton equations for the barred variables. The remaining factor $\chi$ comes from the change of propagation variable $z=\chi\xi$.
We thus obtain the following degenerate FWM system that we consider for our projection:
\begin{align}
\bar{H} &= \frac{\tilde{H}}{2 \chi^{3}} = - \frac{\varsigma \chi }{4}\left( \bar{u}_1 \bar{u}_2 \bar{u}^{2}_3 + \bar{v}_1 \bar{v}_2 \bar{v}^{2}_3 \right) + \frac{\chi \bar{u}^{2}_1 \bar{v}^{2}_1}{4} + \frac{\chi \bar{u}^{2}_2 \bar{v}^{2}_2}{4} + \frac{\chi \bar{u}^{2}_3 \bar{v}^{2}_3}{8} - \frac{1}{\chi}\sum_{j=1}^{3}\bar{u}_j \bar{v}_j, \label{eq:Hbar-deg-fem}
\end{align}
\begin{align}
  \frac{\partial}{\partial z} \bar{u}_j(z) &=  \frac{\partial}{\partial \bar{v}_j} \bar{H}, \notag \\
  \frac{\partial}{\partial z} \bar{v}_j(z) &=  -\frac{\partial}{\partial \bar{u}_j} \bar{H}, \notag
\end{align}
\begin{align}
  \frac{\partial}{\partial z} \bar{u}_1(z) &=  - \frac{\bar{u}_1}{\chi} + \frac{\chi \bar{u}_1^{2} \bar{v}_1}{2} - \frac{\varsigma \chi \bar{v}_2 \bar{v}_3^{2}}{4}, \notag \\
  \frac{\partial}{\partial z} \bar{u}_2(z) &=  - \frac{\bar{u}_2}{\chi} + \frac{\chi \bar{u}_2^{2} \bar{v}_2}{2} - \frac{\varsigma \chi \bar{v}_1 \bar{v}_3^{2}}{4}, \notag \\
  \frac{\partial}{\partial z} \bar{u}_3(z) &=  - \frac{\bar{u}_3}{\chi} + \frac{\chi \bar{u}_3^{2} \bar{v}_3}{4} - \frac{\varsigma \chi \bar{v}_1 \bar{v}_2 \bar{v}_3}{2}, \notag \\
  \frac{\partial}{\partial z} \bar{v}_1(z) &= \frac{\bar{v}_1}{\chi} - \frac{\chi \bar{v}_1^{2} \bar{u}_1}{2} + \frac{\varsigma \chi \bar{u}_2 \bar{u}_3^{2}}{4}, \notag \\
  \frac{\partial}{\partial z} \bar{v}_2(z) &= \frac{\bar{v}_2}{\chi} - \frac{\chi \bar{v}_2^{2} \bar{u}_2}{2} + \frac{\varsigma \chi \bar{u}_1 \bar{u}_3^{2}}{4}, \notag \\
  \frac{\partial}{\partial z} \bar{v}_3(z) &= \frac{\bar{v}_3}{\chi} - \frac{\chi \bar{v}_3^{2} \bar{u}_3}{4} + \frac{\varsigma \chi \bar{u}_1 \bar{u}_2 \bar{u}_3}{2}, \label{eq:ubar-vbar-degfwm-system}
\end{align}
There is a projection from \eqref{eq:ubar-vbar-degfwm-system} to the hatted normal form \eqref{eq:uv-hat-system}, which we now describe.

Let us define modes $\hat{u}_j$, $\hat{v}_j$ with $j=1,2$ (note the hat notation), in terms of $\bar{u}_k$, $\bar{v}_k$ modes with $k=1,2,3$, as follows:
\begin{align}
  \hat{u}_1{\left(z\right)} &= \frac{\sqrt{2} \sqrt{{\gamma}_{1} - {\lambda}_{1}} \bar{u}_1{\left(z\right)} e^{z \theta}}{\bar{v}_3{\left(z\right)}}, \notag \\
  \hat{u}_2{\left(z\right)} &= \frac{\sqrt{2} \sqrt{{\gamma}_{2} - {\lambda}_{1}} \bar{u}_2{\left(z\right)} e^{-z \theta}}{\bar{v}_3{\left(z\right)}}, \notag \\
  \hat{v}_1{\left(z\right)} &= \frac{\sqrt{2} \sqrt{{\gamma}_{1} - {\lambda}_{1}} \bar{v}_1{\left(z\right)} e^{-z \theta}}{\bar{u}_3{\left(z\right)}}, \notag \\
  \hat{v}_2{\left(z\right)} &= \frac{\sqrt{2} \sqrt{{\gamma}_{2} - {\lambda}_{1}} \bar{v}_2{\left(z\right)} e^{z \theta}}{\bar{u}_3{\left(z\right)}}, \label{eq:projection-hat-as-bars}
\end{align}
where:
\begin{equation}
\theta = - {b}_{2} {\gamma}_{1} + \frac{2 {\gamma}_{1}}{{b}_{0} + {b}_{1} {\lambda}_{1} + {b}_{2} {\lambda}_{1}^{2}},
\end{equation}
and where $\gamma_1, \gamma_2, b_0, b_1, b_2$ all retain their definitions given in Appendices \ref{app:param-def-appendix} and \ref{app:init-conds} for the two mode coherent coupler system, and $\lambda_1$ remains the chosen root of \eqref{eq:lambda-root}.
Let us then set the values of the conserved quantities in the degenerate FWM system (bar coordinates) in terms of parameters from the coherent coupler system (hat coordinates). 
To do this, we first define the constant $d_5$ in terms of $\lambda_1$ and the other $d_j$ parameters from the coherent coupler system given in Appendix \ref{app:param-def-appendix} such that:
\begin{equation}
{d}_{5} = {d}_{1} + 2 {d}_{2} {\lambda}_{1} + 3 {d}_{3} {\lambda}_{1}^{2} + 4 {d}_{4} {\lambda}_{1}^{3},
\end{equation}
then we fix the following relationship between the intermodal power constants $\bar{\gamma}_j$ in the degenerate FWM system and the $\gamma_k$ of the coherent coupler system such that:
\begin{alignat}{2}
\bar{u}_1\bar{v}_1 - \bar{u}_2\bar{v}_2 &= \bar{\gamma}_1 - \bar{\gamma}_2 &&= \frac{{d}_{5} {\gamma}_{1}}{4 \left({\gamma}_{1}^{2} -  {\lambda}_{1}^{2}\right)}, \notag \\
2\bar{u}_1\bar{v}_1 - \bar{u}_3\bar{v}_3 &= 2\bar{\gamma}_1 - \bar{\gamma}_3 &&= \frac{d_5}{4\left(\gamma_1 - \lambda_1\right)}, \notag \\
2\bar{u}_2\bar{v}_2 - \bar{u}_3\bar{v}_3 &= 2\bar{\gamma}_2 - \bar{\gamma}_3 &&= \frac{d_5}{4\left(\gamma_2 - \lambda_1\right)},
\end{alignat}
where, by definition:
\begin{equation}
\sum_{j=1}^{3} {\bar{\gamma}}_{j} = \sum_{j=1}^{2} {\gamma}_{j} = 0.
\end{equation}
We then fix the free parameter $\chi$ and the sign parameter $\varsigma=\pm1$ in the degenerate FWM system in terms of parameters from the coherent coupler system such that:
\begin{align}
  \chi &= \frac{8 \left( {b}_{0} + {b}_{1} {\lambda}_{1} + {b}_{2} {\lambda}_{1}^{2} \right)}{{d}_{5}}, \\
  \varsigma &= \frac{2 \sqrt{{\gamma}_{2} - {\lambda}_{1}} \sqrt{{\gamma}_{1} - {\lambda}_{1}}}{{b}_{0} + {b}_{1} {\lambda}_{1} + {b}_{2} {\lambda}_{1}^{2}},
\end{align}
and the Hamiltonian in the degenerate FWM system takes the value:
\begin{equation}
  \bar{H} = - \frac{{b}_{2} {d}_{5}}{8} - \frac{2}{\chi} + \frac{64 \left({d}_{5} + 16 {\lambda}_{1}\right) {\lambda}_{1}}{\chi^{3} {d}_{5}^{2}}.
\end{equation}
These formulae define the projection on a generic chart with $d_5\ne0$, $Q(\lambda_1)\ne0$, $\gamma_j\ne\lambda_1$, and $\bar u_3\bar v_3\ne0$.
Square roots must be chosen compatibly with the displayed value of $\varsigma$; its square is one because the root equation gives $Q(\lambda_1)^2=4(\gamma_1-\lambda_1)(\gamma_2-\lambda_1)$.
The three displayed power differences in the barred system contain only two independent relations.
With these parameter relationships the projection is defined. Differentiating \eqref{eq:projection-hat-as-bars} with respect to $z$, substituting \eqref{eq:ubar-vbar-degfwm-system} into the right-hand side, replacing ratios of bar-coordinate modes by hat-coordinate modes using \eqref{eq:projection-hat-as-bars}, and simplifying using the Hamiltonian in \eqref{eq:Hbar-deg-fem}, recovers the two-mode coherent coupler system in \eqref{eq:uv-hat-system}.
This establishes a projection from degenerate FWM to the coherent coupler. 

The solutions to \eqref{eq:ubar-vbar-degfwm-system} are for $j=1,2,3$:
\begin{align}
\bar{u}_j{\left(z\right)} &= \sqrt{\frac{s\left(j\right)}{2}}\frac{{\bar{\alpha}}_{j}\sigma{\left(z - 2 z_{0} + {\mu}_{j}\right)} e^{z \left(\frac{\chi {\bar{\gamma}}_{j}}{s{\left(j \right)}} - \zeta{\left({\mu}_{j} - z_{0}\right)} + \frac{1}{\chi}\right)}}{\sigma{\left({\mu}_{j} - z_{0}\right)} \sigma{\left(z - z_{0}\right)  }}, \notag \\
\bar{v}_j{\left(z\right)} &= \sqrt{\frac{s\left(j\right)}{2}}\frac{\sigma{\left(z - {\mu}_{j}\right)} e^{- z \left(\frac{\chi {\bar{\gamma}}_{j}}{s{\left(j \right)}} - \zeta{\left({\mu}_{j} - z_{0}\right)} + \frac{1}{\chi}\right)}}{{\bar{\alpha}}_{j} \sigma{\left({\mu}_{j} - z_{0}\right)} \sigma{\left(z - z_{0}\right)} }, \notag \\
\bar{u}_j{\left(z\right)} \bar{v}_j{\left(z\right)} &= \frac{s{\left(j \right)}}{2}\left(\wp{\left({\mu}_{j} - z_{0} \right)} - \wp{\left(z - z_{0}\right)}\right) \label{eq:-ubar-vbar-sols-dfwm}
\end{align}
where $\bar{\alpha}_j$ is an integration constant that captures initial relative phase offsets between modes and can be used to equate initial conditions between the degenerate FWM and coherent coupler systems. 
The function $s(j)$ is the degeneracy weight, with $s(1)=s(2)=1$ and $s(3)=2$, and the new third $\mu_j$ constant is $\mu_3 = z_1 + z_0$. The elliptic functions use invariants $g_2, g_3$, and $\mu_1, \mu_2, z_0, z_1, g_2, g_3$ share the same definitions and values as those given in Section~\ref{sec:mode-power} for the coherent coupler, thereby connecting the two systems. 
The solutions in \eqref{eq:-ubar-vbar-sols-dfwm} are exponentially gauged Kronecker kernels; their Baker--Akhiezer normalisation is made explicit in Section~\ref{sec:discussion}.
These functions are named after Kronecker, who developed both Fourier series \cite{whittakerCourseModernAnalysis2021} and multipole-expansion \cite{weilEllipticFunctionsAccording1976, charolloisEllipticFunctionsAccording2016} representations that may prove useful for further analysis in nonlinear optics.
Importantly, while the $\sigma$ functions and exponentials are entire functions, their ratio is a single-valued meromorphic function with well-defined poles and zeros, in contrast to the multi-valued solutions in \eqref{eq:u-v-quartic}.
The solutions are quasi-periodic, and modulo the period lattice they have zeros where the numerator $\sigma$ function vanishes and poles where the denominator $\sigma$ function vanishes.

The Kronecker theta form establishes a direct connection with other integrable nonlinear optical systems, including two-wave and three-wave mixing in quadratic nonlinear media, polarisation dynamics in nonlinear fibres, and parity-time symmetric nonlinear couplers \cite{heskethGeneralComplexEnvelope2015}, suggesting a unified mathematical framework across different parametric processes.

% ---------- SECTION -------------
% --------------------------------

\section{The Frobenius-Stickelberger determinant}
\label{app:fs-det-ham-proof}

The Frobenius-Stickelberger (FS) determinant formula is an elliptic function identity relating products of $\sigma$ functions to determinants of derivatives of $\wp$ \cite{whittakerCourseModernAnalysis2021, frobeniusZurTheorieElliptischen1877}.
We represent the conserved Hamiltonian product relation by this identity. The argument uses Hamiltonian conservation; it does not derive that conservation independently from the determinant.

Our starting point is equation \eqref{eq:fs-det-start}, which relates the wave mixing product to the phase modulation terms via the Hamiltonian.
We show that this equation can be transformed into the FS determinant formula by treating the left-hand side (the product of modes) and right-hand side (the polynomial in $\rho$) separately, each ultimately expressed in terms of Weierstrass functions.

We are interested in the $4\times4$ case of FS in the form:
\begin{align}
\label{eq:fs-det-m}
\frac{C\,\sigma{\left(z + {\nu}_{1}\right)} \sigma{\left(z + {\nu}_{2}\right)} \sigma{\left(z + {\nu}_{3}\right)} \sigma{\left(z - {\nu}_{1} - {\nu}_{2} - {\nu}_{3}\right)}}{\sigma^{4}{\left(z\right)}} & = \det M
\end{align}
where:
\begin{align}
C = &\frac{12  \sigma{\left({\nu}_{1} - {\nu}_{2}\right)} \sigma{\left({\nu}_{1} - {\nu}_{3}\right)} \sigma{\left({\nu}_{2} - {\nu}_{3}\right)} }{\sigma^{4}{\left({\nu}_{1}\right)} \sigma^{4}{\left({\nu}_{2}\right)} \sigma^{4}{\left({\nu}_{3}\right)}} \\[6pt]
\det M = &
\begin{vmatrix}
1 & \wp(z) & \frac{\partial \wp(z)}{\partial z} & \frac{\partial^2 \wp(z)}{\partial z^2} \\
1 & \wp(\nu_1) & -\frac{\partial \wp(\nu_1)}{\partial \nu_1} & \frac{\partial^2 \wp(\nu_1)}{\partial \nu_1^2} \\
1 & \wp(\nu_2) & -\frac{\partial \wp(\nu_2)}{\partial \nu_2} & \frac{\partial^2 \wp(\nu_2)}{\partial \nu_2^2} \\
1 & \wp(\nu_3) & -\frac{\partial \wp(\nu_3)}{\partial \nu_3} & \frac{\partial^2 \wp(\nu_3)}{\partial \nu_3^2} \\
\end{vmatrix}  \notag \\[6pt]
= & - 6 \left(\wp{\left({\nu}_{1}\right)} - \wp{\left(z\right)}\right) \left(\wp{\left({\nu}_{2}\right)} - \wp{\left(z\right)}\right) \left(\wp{\left({\nu}_{1}\right)} - \wp{\left({\nu}_{2}\right)}\right) \wp'{\left({\nu}_{3}\right)} \notag \\
& + 6 \left(\wp{\left({\nu}_{1}\right)} - \wp{\left(z\right)}\right) \left(\wp{\left({\nu}_{3}\right)} - \wp{\left(z\right)}\right) \left(\wp{\left({\nu}_{1}\right)} - \wp{\left({\nu}_{3}\right)}\right) \wp'{\left({\nu}_{2}\right)}  \notag \\
& - 6 \left(\wp{\left({\nu}_{2}\right)} - \wp{\left(z\right)}\right) \left(\wp{\left({\nu}_{3}\right)} - \wp{\left(z\right)}\right) \left(\wp{\left({\nu}_{2}\right)} - \wp{\left({\nu}_{3}\right)}\right) \wp'{\left({\nu}_{1}\right)}  \notag \\
& - 6 \left(\wp{\left({\nu}_{1}\right)} - \wp{\left({\nu}_{2}\right)}\right) \left(\wp{\left({\nu}_{1}\right)} - \wp{\left({\nu}_{3}\right)}\right) \left(\wp{\left({\nu}_{2}\right)} - \wp{\left({\nu}_{3}\right)}\right) \wp'{\left(z\right)}
\end{align}

The starting point for us is the following relation established via \eqref{eq:ham-uv}, \eqref{eq:duv_j}, and \eqref{eq:Q-uv}, where any choice of $j$ may be used in the modal-power derivative:
\begin{align}
2 \prod_{k=1}^{2} u_k &= {a}_{0} + \sum_{k=1}^{2} {a}_{k} u_k v_k  + \frac{1}{2}\sum\limits_{k,l=1}^{2} {a}_{k,l} u_k v_k u_l v_l - \frac{\partial}{\partial z}(u_j v_j) \notag \\
&=\rho' + \sum_{l=0}^{2}b_l\,\rho\left(z\right)^l. \label{eq:fs-det-start}
\end{align}
We now transform \eqref{eq:fs-det-start} into \eqref{eq:fs-det-m}.

\subsection{The left-hand side}
From \eqref{eq:uv-wp}, the right-hand side of \eqref{eq:fs-det-start} is a doubly periodic elliptic function, and therefore so is the left-hand side. Choose primitive lattice generators $2\omega_1,2\omega_3$ with $\operatorname{Im}(\omega_3/\omega_1)>0$, and write $\eta_j=\zeta(\omega_j)$ for $j=1,3$.
For integers $m,n$ we have:
\begin{align}
\prod_{j=1}^{2} \frac{u_j{\left(z\right)}}{u_j{\left(2 m {\omega}_{3} + 2 n {\omega}_{1} + z\right)}} & = 1. \label{eq:u-prod-is-one}
\end{align}
We substitute our solution \eqref{eq:u-v-quartic} into \eqref{eq:u-prod-is-one}, take advantage of \eqref{eq:r1j-sum} to simplify, and use the quasi-periodicity of $\sigma$:
\begin{align}
\sigma{\left(2 m {\omega}_{3} + 2 n {\omega}_{1} + z\right)} = \left(-1\right)^{m n + m + n} \sigma{\left(z\right)} e^{\left(2 m {\omega}_{3} + 2 n {\omega}_{1} + 2 z\right) \left(m\eta_3+n\eta_1\right)} \label{eq:quasi-sigma}
\end{align}
to reduce the product in \eqref{eq:u-prod-is-one} to an exponential which, as it is equal to one, must have argument $2\pi i \mathrm{N}(n,m)$, with $\mathrm{N}(n,m)$ an integer that varies with $n,m$.
Put $\nu_j=\mu_j-z_0$, $S=2z_1+\nu_1+\nu_2$ and $R=r_{0,1}+r_{0,2}$.
The product multiplier gives
\begin{equation}
-2(m\eta_3+n\eta_1)S-(2m\omega_3+2n\omega_1)R=2\pi i\,\mathrm N(n,m).
\label{eq:exp-arg-is-one}
\end{equation}
Using $\eta_1\omega_3-\eta_3\omega_1=i\pi/2$ for the two primitive shifts gives integers $p,q$ such that
\begin{align}
2z_1+\nu_1+\nu_2&=2p\omega_1+2q\omega_3,\label{eq:nuj-sum-zero}\\
\sum_{j=1}^2 r_{0,j}&=-2p\eta_1-2q\eta_3.\label{eq:roj-is-zeta}
\end{align}
These expressions use quasi-period constants, not zeta evaluated at a possibly singular lattice point.
Ultimately, equations \eqref{eq:quasi-sigma}, \eqref{eq:nuj-sum-zero} and \eqref{eq:roj-is-zeta} allow us to make the following substitution (or the equivalent at $z\rightarrow z - z_0$) in the left-hand side of \eqref{eq:fs-det-start}:
\begin{align}
\frac{\sigma{\left(z + z_1 \right)}}{\sigma{\left(z_1 \right)}}\,\exp{\left(z \sum\limits_{j=1}^{2} {r}_{0,j}\right)} = - \frac{\sigma{\left(z - z_1 - {\nu}_{1} - {\nu}_{2}\right)}}{\sigma{\left(z_1 + {\nu}_{1} + {\nu}_{2}\right)}}.
\end{align}

\subsection{The right-hand side}
The following polynomial interpolation formula holds for any quadratic polynomial:
\begin{align}
\sum_{l=0}^{2}b_l\,\rho\left(z\right)^l =\, &\frac{\rho{\left(z \right)} - {\gamma}_{2}}{{\gamma}_{1} - {\gamma}_{2}} \sum\limits_{l=0}^{2}b_l\,\gamma_1^l + \notag \\
&\frac{\rho{\left(z \right)} - {\gamma}_{1}}{{\gamma}_{2} - {\gamma}_{1}} \sum\limits_{l=0}^{2}b_l\,\gamma_2^l + \notag \\
&\left({\gamma}_{1} - \rho{\left(z \right)}\right) \left({\gamma}_{2} - \rho{\left(z \right)}\right) b_2 \label{eq:poly-interp}
\end{align}
where, in our case, $\gamma_j=\rho(\mu_j)$ and the sum over $b_l \gamma_j^l$ has the following sign fixed by the derivative sheet in \eqref{eq:wp-points}:
\begin{align}
\sum\limits_{l=0}^{2}b_l\,\gamma_j^l = \rho'\left(\mu_j\right).
\end{align}
We can then use \eqref{eq:uv-wp} to express $\rho$ in terms of $\wp$ so that all variables in \eqref{eq:poly-interp} can be written in terms of $\wp$ and $\wp'$. 
For an explicit determinant comparison, translate $z\mapsto z+z_0$, so that the original argument $z-z_0$ becomes $z$, and put $\nu_3=z_1$.
Divide the translated relation \eqref{eq:fs-det-start} by
\begin{equation}
B(z)=\frac{\wp'(z_1)}{b_2[\wp(z_1)-\wp(z)]^2}.
\end{equation}
The reduction gives $\rho'/B=\wp'(z)$, while $Q(\rho)/B$ is a quadratic polynomial in $\wp(z)$.
At $\wp(z)=\wp(\nu_j)$ its values are $\wp'(\nu_j)$ for $j=1,2$ by the selected sheets, and its value at $\wp(z_1)$ is $\wp'(z_1)$ because the leading coefficient of $Q$ is $b_2$.
Consequently, with $w_j=\wp(\nu_j)$, $y_j=\wp'(\nu_j)$ and $w=\wp(z)$,
\begin{equation}
\frac{\rho'+Q(\rho)}{B}=\wp'(z)+L(w),\qquad
L(w)=\sum_{j=1}^3 y_j\prod_{\substack{k=1\\k\ne j}}^3\frac{w-w_k}{w_j-w_k}.
\end{equation}
Expanding the determinant in \eqref{eq:fs-det-m}, using $\wp''=6\wp^2-g_2/2$, gives
\begin{equation}
\det M=-6(w_1-w_2)(w_1-w_3)(w_2-w_3)[\wp'(z)+L(w)].
\end{equation}
The translated mode product, after the exponential absorption above and division by $B$, is proportional to the sigma ratio in \eqref{eq:fs-det-m}.
Multiplication by $\sigma(z)^4$ and comparison at $z=0$ fixes the proportionality constant and gives $C$ as displayed.
The calculation assumes distinct interpolation nodes and nonzero denominators; coincident cases require limits where these exist.
Thus the Hamiltonian product relation is represented by the four-point Frobenius--Stickelberger identity in the elliptic solution coordinates.

\section{Numerical validation and implementation} \label{sec:num-val}

In this section we validate the analytic solutions through numerical evaluation and demonstrate their practical implementation using open-source software.
This serves two purposes: first, to verify that the solutions correctly satisfy the differential equations, and second, to show that the Weierstrass elliptic functions can be readily evaluated using standard numerical libraries.

We evaluate the analytic solutions using the \emph{pyweierstrass} Python package \cite{laurentPyweierstrass2022}, which provides Weierstrass elliptic functions as a wrapper around Jacobi theta functions from the \emph{mpmath} package \cite{MpmathPythonLibrary2023}. 
For comparison, we solve the differential equations numerically using the DOP853 Runge-Kutta algorithm from \emph{SciPy} \cite{virtanenSciPy10Fundamental2020}.

For the independent Hamiltonian variables we plot the modal products and logarithmic phase variables
\begin{align}
\mathcal{P}_j(z) &= u_j(z)v_j(z), \notag \\
\phi_j(z) &= \frac{\log\left(\frac{u_j(z)}{v_j(z)}\right)}{2 i}. \label{eq:A-phi-u-v}
\end{align}

\begin{figure}[H]
\centering
\includegraphics[width=1.0\textwidth]{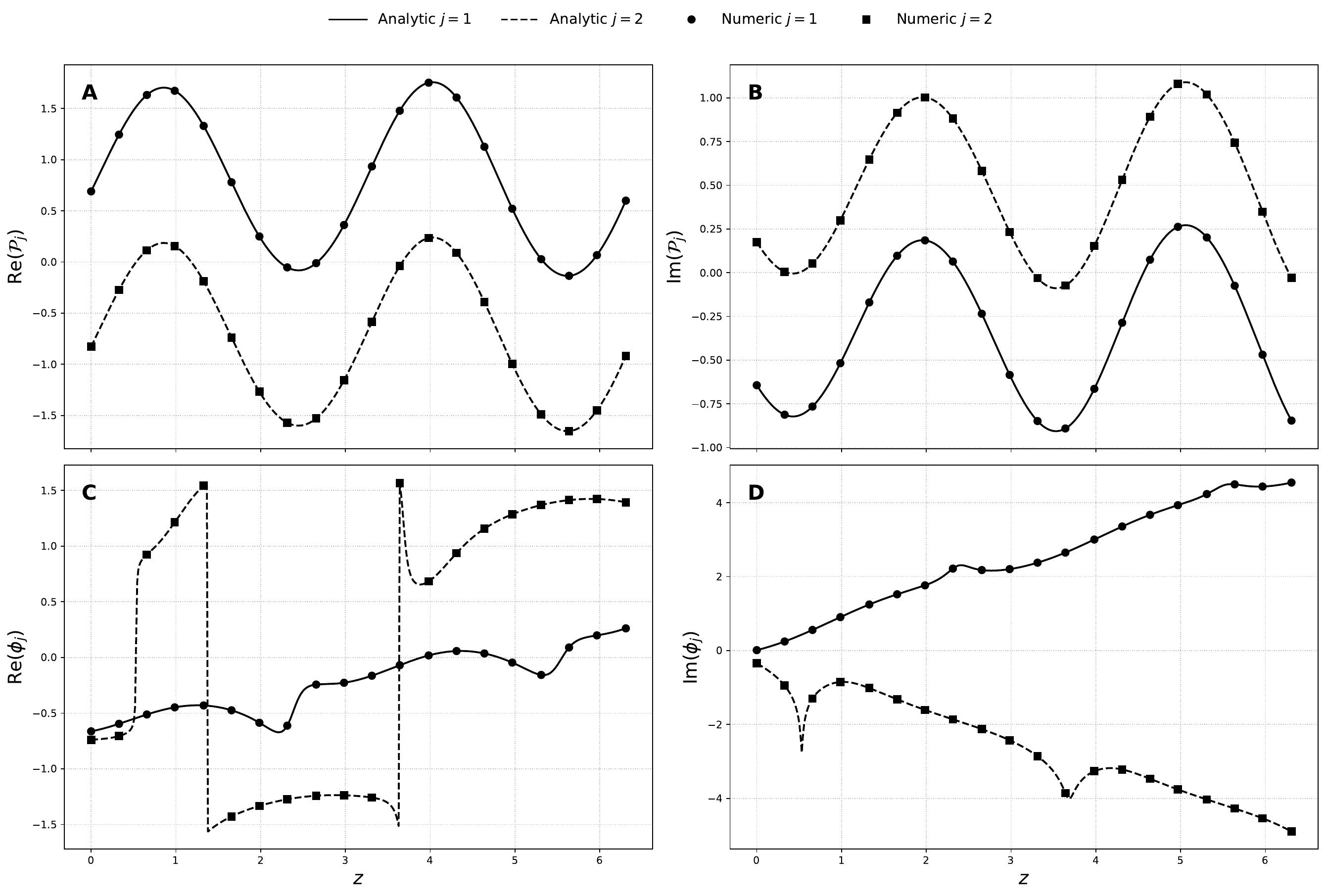}
\caption{\textbf{A} real and \textbf{B} imaginary parts of the complex mode power $\mathcal{P}_j$, and \textbf{C} real and \textbf{D} imaginary parts of the complex phase variable $\phi_j$, using the analytic solutions in \eqref{eq:u-v-quartic} (lines) compared against numerical integration of the original general coherent coupler system in \eqref{eq:uv-system} (symbols).}
\label{fig:case1-plot1}
\end{figure}

\begin{figure}[H]
\centering
\includegraphics[width=1.0\textwidth]{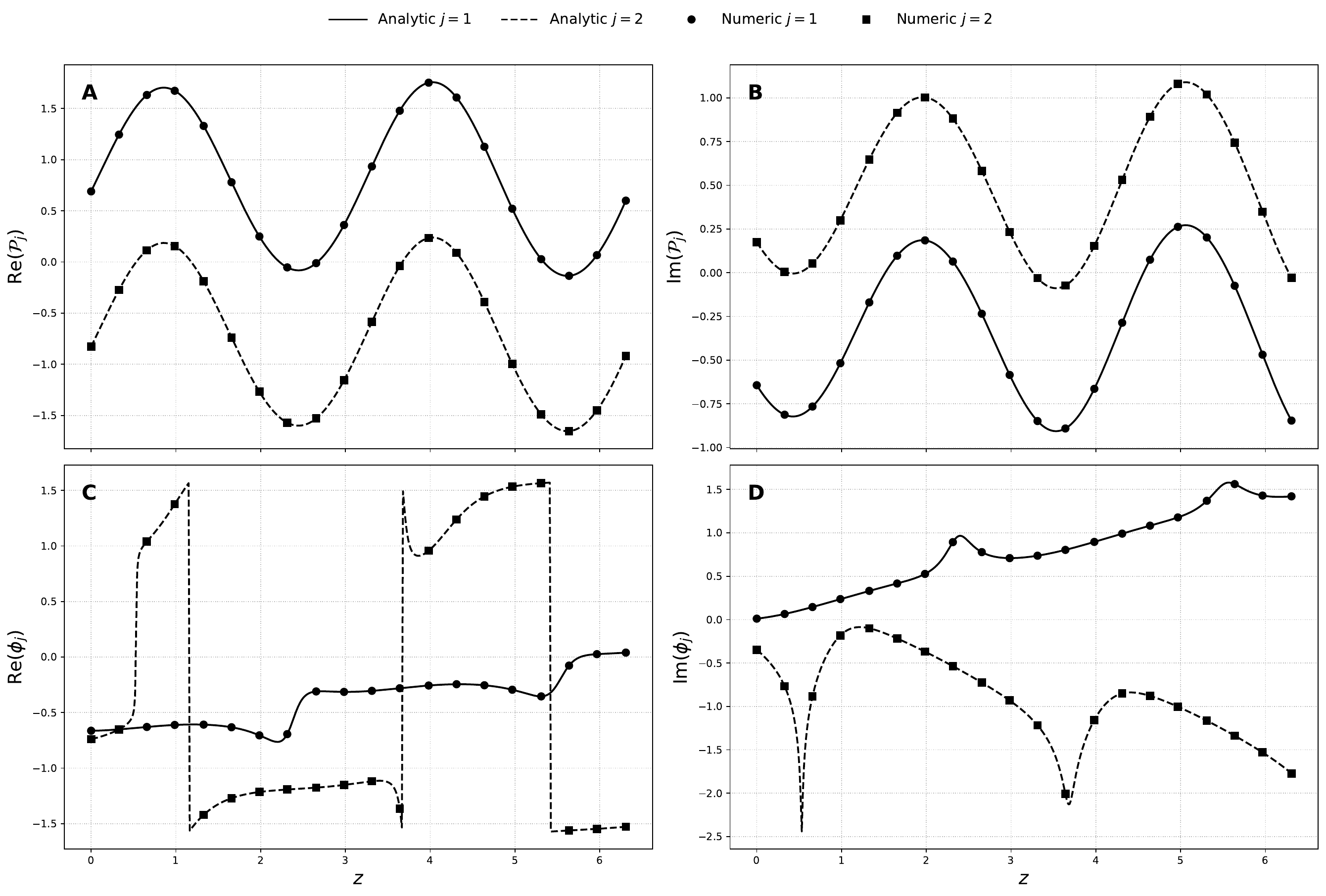}
\caption{\textbf{A} real and \textbf{B} imaginary parts of the complex mode power $\mathcal{P}_j$, and \textbf{C} real and \textbf{D} imaginary parts of the complex phase variable $\phi_j$, using the gauge transformed analytic solutions in \eqref{eq:uhat-vhat-gauge-sols} (lines) compared against numerical integration of the gauge transformed general coherent coupler system in \eqref{eq:uv-hat-system} (symbols).}
\label{fig:case2-plot1}
\end{figure}

\begin{figure}[H]
\centering
\includegraphics[width=1.0\textwidth]{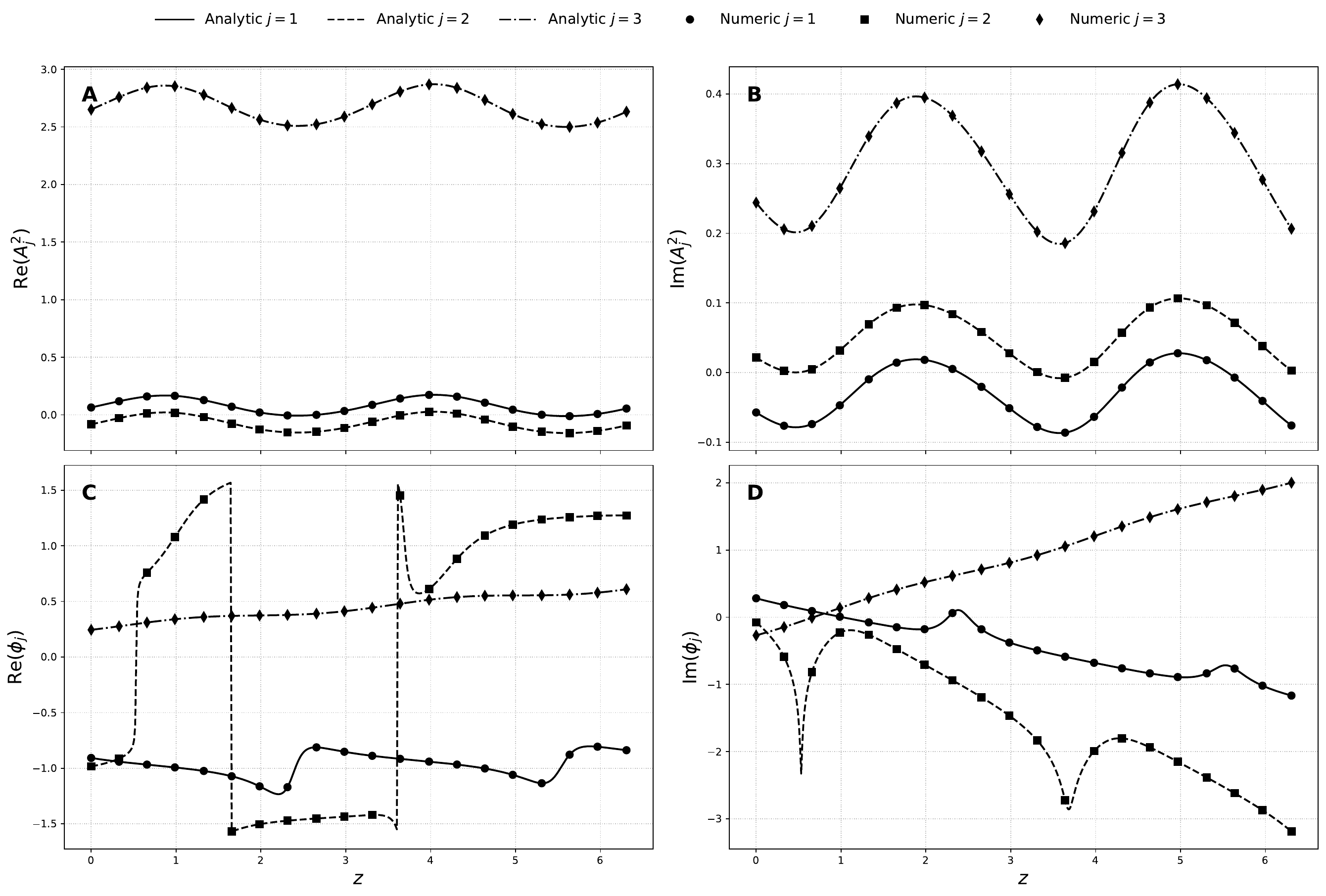}
\caption{\textbf{A} real and \textbf{B} imaginary parts of the complex mode power $\mathcal{P}_j$, and \textbf{C} real and \textbf{D} imaginary parts of the complex phase variable $\phi_j$, using the analytic solutions in \eqref{eq:-ubar-vbar-sols-dfwm} (lines) compared against numerical integration of the three-mode degenerate FWM system in \eqref{eq:ubar-vbar-degfwm-system} (symbols).}
\label{fig:case3-plot1}
\end{figure}

\begin{figure}[H]
\centering
\includegraphics[width=1.0\textwidth]{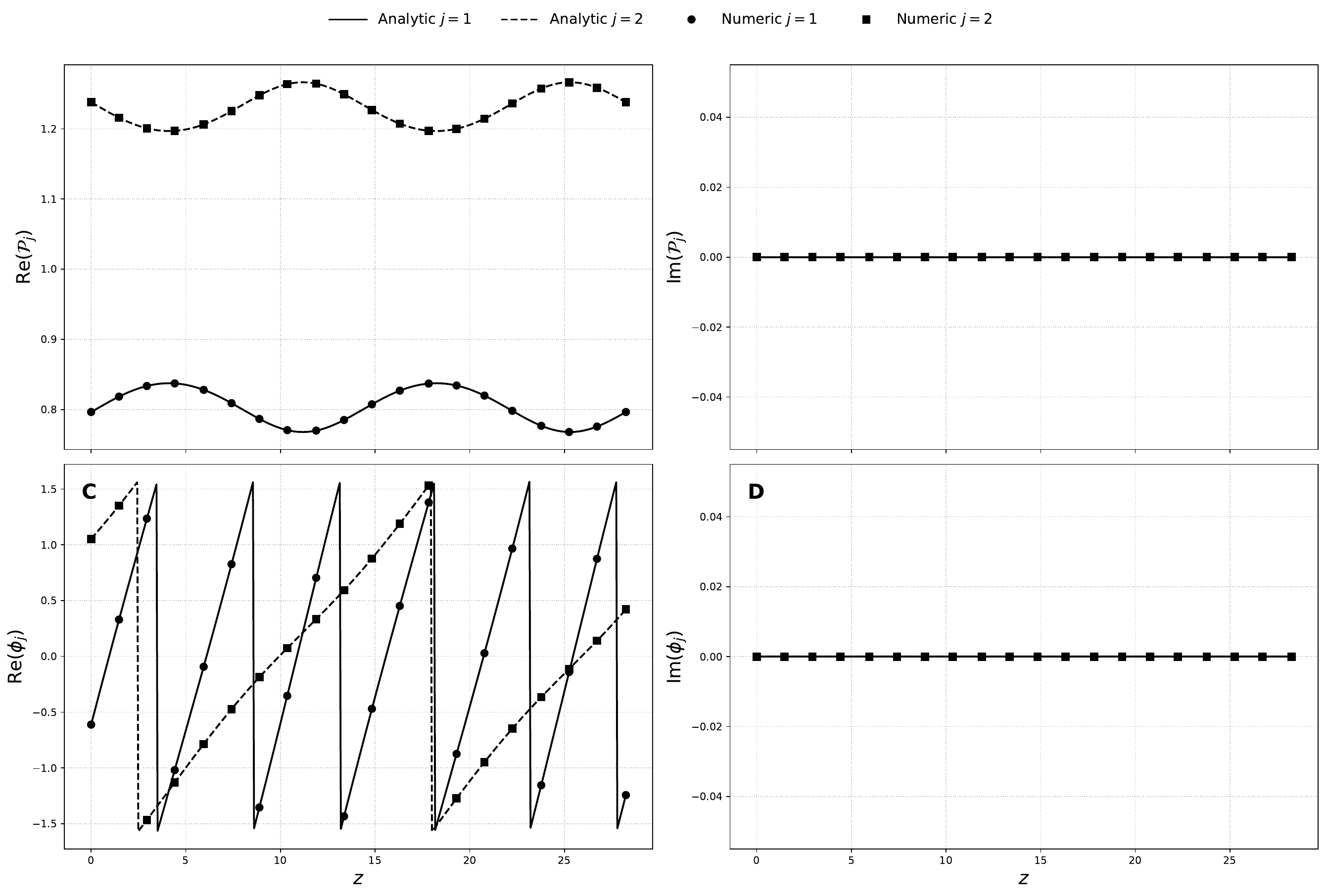}
\caption{\textbf{A} real and \textbf{B} imaginary parts of the mode power $\mathcal{P}_j$, and \textbf{C} real and \textbf{D} imaginary parts of the phase variable $\phi_j$, using the analytic solutions in \eqref{eq:u-v-quartic-no-wp-sqrt} and the relations in \eqref{eq:uv-to-jensen} (lines) compared against numerical integration of Jensen's original system in \eqref{eq:jensen} (symbols).}
\label{fig:case4-plot1}
\end{figure}

These are modal products and logarithmic phase variables, rather than optical powers and phases in general.
On the physical slice \eqref{eq:physical-slice}, the optical powers are $-\mathcal P_1$ and $\mathcal P_2$; optical phases are recovered using the field map, including its conjugation and constant phase factors.
In Figures~\ref{fig:case1-plot1}--\ref{fig:case4-plot1}, jumps in the logarithmic phase diagnostic reflect its principal-branch convention. This plotting convention is separate from the continuous logarithm required in the envelope reconstruction.

The top row of Figure \ref{fig:case1-plot1} shows the real, \textbf{A}, and imaginary, \textbf{B}, parts of the complex mode power $\mathcal{P}_j$,  
while the bottom row shows the real, \textbf{C}, and imaginary, \textbf{D}, parts of the complex phase variable $\phi_j$, using the analytic solutions in \eqref{eq:u-v-quartic} compared against numerical integration of the original general coherent coupler system in \eqref{eq:uv-system}.
The intermodal power conservation laws in \eqref{eq:power-conserved} are evident from the coordinated evolution of the two modes. 
The initial mode and parameter values for this plot were $u_1(0)=0.487-0.828i$, $u_2(0)=-0.971-0.865i$, $v_1(0)=0.942 + 0.279i$, $v_2(0)=0.386-0.523i$, $a_1=-0.161$, $a_2=-0.897$, $a_{1,1}=0.583$, $a_{1,2}=a_{2,1}=-0.0754$, and $a_{2,2}=-0.0525$.

The top row of Figure \ref{fig:case2-plot1} shows the real, \textbf{A}, and imaginary, \textbf{B}, parts of the complex mode power $\mathcal{P}_j$,  
while the bottom row shows the real, \textbf{C}, and imaginary, \textbf{D}, parts of the complex phase variable $\phi_j$, using the gauge transformed analytic solutions in \eqref{eq:uhat-vhat-gauge-sols} compared against numerical integration of the gauge transformed general coherent coupler system in \eqref{eq:uv-hat-system}.
Here, $\hat{u}_j, \hat{v}_j$ are used in place of $u_j, v_j$ in \eqref{eq:A-phi-u-v}.
The initial values of $\hat{u}_j, \hat{v}_j$ were set equal to the values of $u_j, v_j$ used in Figure \ref{fig:case1-plot1} and the $a_1$, $a_2$, $a_{1,1}$, $a_{1,2}$, $a_{2,1}$, and $a_{2,2}$ parameters were the same in all figures.
In comparison with Figure \ref{fig:case1-plot1}, the modal power $\mathcal{P}_j$ is unchanged after the gauge transform but the phase $\phi_j$ changes as expected from the definition of the transform.

The top row of Figure \ref{fig:case3-plot1} shows the real, \textbf{A}, and imaginary, \textbf{B}, parts of the complex mode power $\mathcal{P}_j$,  
while the bottom row shows the real, \textbf{C}, and imaginary, \textbf{D}, parts of the complex phase variable $\phi_j$, using the analytic solutions in \eqref{eq:-ubar-vbar-sols-dfwm} compared against numerical integration of the three-mode degenerate FWM system in \eqref{eq:ubar-vbar-degfwm-system}.
Here, $\bar{u}_j, \bar{v}_j$ are used in place of $u_j, v_j$ in \eqref{eq:A-phi-u-v}.
The corresponding intermodal power conservation laws are evident from the coordinated evolution of the three modes.
We set $\bar{u}_3(0)=\sigma(z_0 - z_1)/\left(\sigma(z_0)\sigma(z_1)\right)$, and the remaining initial conditions for the other modes in the bar coordinates are then determined using \eqref{eq:projection-hat-as-bars}, where the values for $\hat{u}_j(0), \hat{v}_j(0)$ and all other parameters share the same values as those used in Figure \ref{fig:case2-plot1}.

Finally, the top row of Figure \ref{fig:case4-plot1} shows the real, \textbf{A}, and imaginary, \textbf{B}, parts of the mode power $\mathcal{P}_j$,  
while the bottom row shows the real, \textbf{C}, and imaginary, \textbf{D}, parts of the phase variable $\phi_j$, using the analytic solutions in \eqref{eq:u-v-quartic-no-wp-sqrt} and the relations in \eqref{eq:uv-to-jensen}, compared against numerical integration of Jensen's original system in \eqref{eq:jensen}.
Here, $A_j, A^{*}_j$ are used in place of $u_j, v_j$ in \eqref{eq:A-phi-u-v} and thus in this case those variables reduce to absolute magnitude squared and the genuine phase, i.e., both become real variables as confirmed by no imaginary parts observed in Figure \ref{fig:case4-plot1} \textbf{B} and \textbf{D}.
The initial mode and parameter values for this plot were $A_1(0)=0.731-0.513i$, $A_2(0)=0.551+0.967i$, $\Omega_1=0.462$, $\Omega_2=-0.00783$, $\Omega_3=-0.541$, and $\Omega_4=0.252$.

In Figures~\ref{fig:case1-plot1}--\ref{fig:case4-plot1} the analytic solutions (lines) are in excellent agreement with the numerical solutions (symbols).
For the parameter sets shown in Figures~\ref{fig:case1-plot1}--\ref{fig:case4-plot1}, the maximum pointwise absolute discrepancy between the analytic solution vectors and the DOP853 numerical solutions, over the plotted interval and sampled points, was approximately $10^{-13}$, $10^{-13}$, $10^{-14}$, and $10^{-15}$, respectively.
These comparisons confirm both the correctness of the analytic solutions and the feasibility of their numerical evaluation using readily available software tools.

\subsection{An asymmetric conservative physical example}

\begin{figure}[htbp]
\centering
\includegraphics[width=\textwidth]{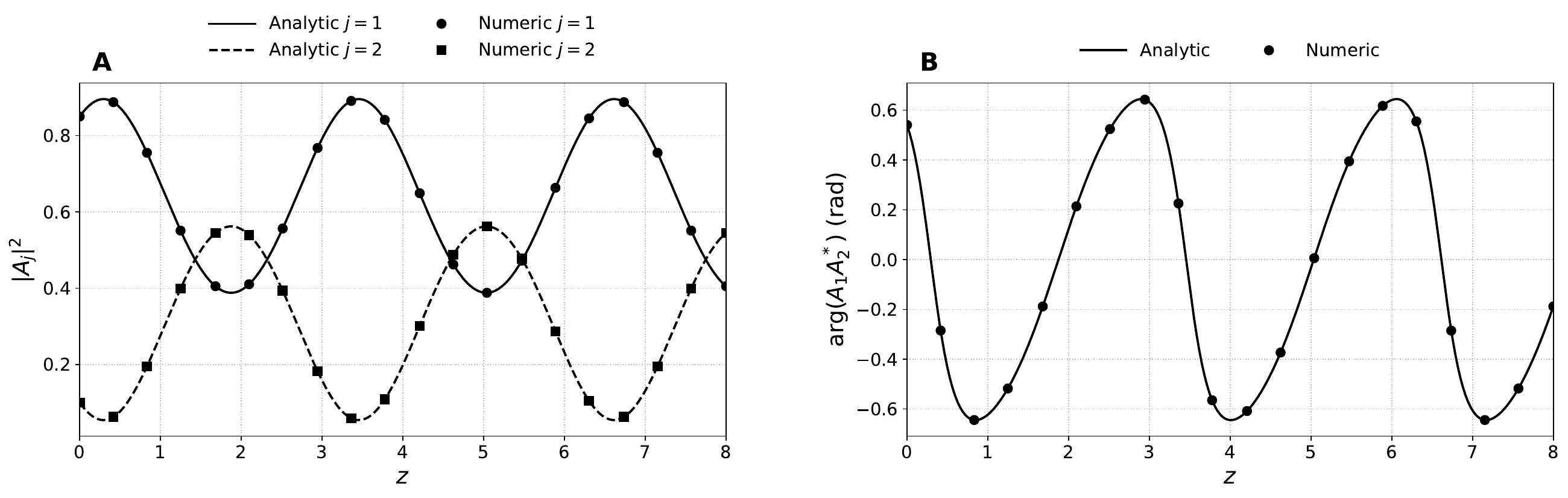}
\caption{An asymmetric conservative coupler: optical powers (left) and continuously unwrapped relative phase (right). Solid and dashed lines show the elliptic powers for modes 1 and 2; circles and squares show the corresponding direct numerical solutions. The relative phase uses a solid line and circles. Distance and power are normalised.}
\label{fig:asymmetric-physical}
\end{figure}

To test unequal coefficients on the physical slice, consider the normalised field equations
\begin{equation}
\frac{dA_j}{dz}=i\left(\beta_j+\sum_{k=1}^2G_{j,k}|A_k|^2\right)A_j+iA_{3-j},\qquad j=1,2,
\label{eq:asymmetric-physical}
\end{equation}
with $\beta_1=0.3$, $\beta_2=-0.2$, $G_{1,1}=1.2$, $G_{2,2}=0.7$, and $G_{1,2}=G_{2,1}=0.25$.
The fields are mapped as in \eqref{eq:uv-to-jensen}, with $a_1=-i\beta_1$, $a_2=i\beta_2$, $a_{j,j}=iG_{j,j}$ and $a_{1,2}=-iG_{1,2}$.
We take $A_1(0)=0.9+0.2i$ and $A_2(0)=0.3-0.1i$.
Figure~\ref{fig:asymmetric-physical} compares the reconstructed optical powers and relative phase with direct integration of \eqref{eq:asymmetric-physical} over $0\le z\le8$.
The maximum absolute error in either complex field at the 501 sampled points is $7.7\times10^{-13}$.
This example tests the generic unequal-self-phase case in a conservative physical system, with total power $0.95$ and both conjugation conditions preserved.

\FloatBarrier
\section{Discussion}
\label{sec:discussion}

The meromorphic modes in the degenerate FWM lift admit two closely related function descriptions.
Let $\vartheta(t;\tau)$ be the odd Jacobi theta function with quasiperiods $1,\tau$, and define the Kronecker kernel \cite[Section~3.4]{brownMultipleEllipticPolylogarithms2013} by
\begin{equation}
F(s,t;\tau)=\frac{\vartheta'(0;\tau)\vartheta(s+t;\tau)}{\vartheta(s;\tau)\vartheta(t;\tau)}.
\end{equation}
With $\tau=\omega_3/\omega_1$ and $\eta_1=\zeta(\omega_1)$, its Baker--Akhiezer normalisation is
\begin{align}
\Phi(x,a)&=\frac{\sigma(x+a)}{\sigma(x)\sigma(a)}e^{-x\zeta(a)}\notag\\
&=\frac{1}{2\omega_1}e^{x[\eta_1a/\omega_1-\zeta(a)]}
F\!\left(\frac{x}{2\omega_1},\frac{a}{2\omega_1};\tau\right).
\label{eq:ba-kernel}
\end{align}
This is the genus-one Baker--Akhiezer function, with the argument convention $a\mapsto-a$ relative to \cite[Eq.~(1.7)]{kricheverEllipticSolutionsNonlinear1994}.
For fixed $a$ outside the lattice it is meromorphic and quasiperiodic in $x$; as a function of the spectral point $a$ it has prescribed exponential behaviour at the marked point $a=0$.
These meromorphic data and prescribed exponential behaviour are the defining features of the Baker--Akhiezer construction.

To identify the functions in \eqref{eq:-ubar-vbar-sols-dfwm}, put $x=z-z_0$, $\nu_j=\mu_j-z_0$, $c_j=1/\chi+\chi\bar\gamma_j/s(j)$ and $B_j=\bar\alpha_j e^{-z_0\zeta(\nu_j)}$.
Then
\begin{align}
\bar u_j(z)&=\sqrt{\frac{s(j)}2}\,B_j e^{c_jz}\Phi(x,\nu_j),\notag\\
\bar v_j(z)&=-\sqrt{\frac{s(j)}2}\,B_j^{-1}e^{-c_jz}\Phi(x,-\nu_j).
\end{align}
Thus the two names describe kernels related by an explicit exponential normalisation.
The original coupler envelopes are obtained by projection and the reciprocal gauge; their generic branching should be distinguished from the meromorphic functions in this lift.

The lifted equations also give a constant-Wronskian interpretation directly.
Write $s_j=s(j)$, $J_j=\bar u_j\bar v_j$, and define the exponentially gauged modes $\mathcal U_j=e^{-c_jz}\bar u_j$ and $\mathcal V_j=e^{c_jz}\bar v_j$.
Since $\bar\gamma_j=J_j-s_j\sum_kJ_k/4$, substitution of \eqref{eq:ubar-vbar-degfwm-system} gives
\begin{align}
\mathcal W_j&=\mathcal U_j'\mathcal V_j-\mathcal U_j\mathcal V_j'\notag\\
&=s_j\bar H-\frac{4\bar\gamma_j}{\chi}
-\frac{\chi\bar\gamma_j^2}{s_j}
-\frac{\chi s_j}{4}\sum_{k=1}^3\frac{\bar\gamma_k^2}{s_k}.
\label{eq:lifted-wronskian}
\end{align}
The right-hand side consists of conserved quantities, so $\mathcal W_j'=0$.
Differentiating the gauged equations then shows that both $\mathcal U_j$ and $\mathcal V_j$ satisfy the same second-order equation
\begin{align}
\mathcal U_j''&=\mathcal Q_j\mathcal U_j,\qquad
\mathcal V_j''=\mathcal Q_j\mathcal V_j,\notag\\
\mathcal Q_j(z)&=\frac{4}{\chi^2}-\frac{\chi\bar H}{2}
+\frac{\chi^2}{4}\sum_{k=1}^3\frac{\bar\gamma_k^2}{s_k}
+\frac{2\bar\gamma_j}{s_j}-\sum_{k=1}^3J_k(z)\notag\\
&=2\wp(z-z_0)+\wp(\nu_j).
\label{eq:lifted-lame}
\end{align}
The potential is generated polynomially by the modal products.
In elliptic notation this is the one-gap Lam\'e equation \cite[Eq.~(1.9)]{kricheverEllipticSolutionsNonlinear1994}.
This additional exponential gauge identifies the spectral functions without altering the autonomous normal form or the projection used above.

The Weierstrass representation makes these connections visible without suggesting that numerical integration is inadequate for optical modelling.
It provides an explicit setting in which to study the mode equations, their singularities, determinant relations and transformations between optical systems.
The related higher-genus construction \cite{zotero-item-855} replaces elliptic functions with hyperelliptic sigma functions and Abelian integrals, and relates conservation laws to higher-genus determinant identities.
The genus-one modes here provide concrete examples that can guide that extension; a general higher-genus Baker--Akhiezer spectral interpretation requires further analysis.

\section{Conclusion}

We have constructed a nonlinear projection from three-mode degenerate FWM onto an autonomous meromorphic normal form of a generalised conservative coherent coupler.
A reciprocal gauge relates this normal form to the original coupler while preserving its modal products.
Building on established power-transfer and phase descriptions, the explicit Weierstrass envelope representation identifies the branching introduced generically by unequal self-phase coefficients and its removal by the gauge.
This lift identifies the meromorphic modes with exponentially gauged Kronecker or genus-one Baker--Akhiezer kernels, while the Frobenius--Stickelberger determinant represents the Hamiltonian product relation.
Together these results place the two-mode coupler within a broader family of integrable continuous-wave systems.
Numerical comparisons verify the reconstructed fields and transformed systems; the principal contribution is the mathematical structure exposed by these representations.

\section*{Acknowledgements}
The author thanks Maria Dimou for a careful read of the manuscript.

\section*{Code Availability}
The code used in this paper is available at \cite{heskethHeskethGDCoherentcouplerpaper2026} and includes Python Jupyter notebooks with symbolic derivations using \emph{SymPy} \cite{meurerSymPySymbolicComputing2017} as well as the code for numerical validation used in Section~\ref{sec:num-val}.

\appendix

% ---------- APPENDIX ------------
% --------------------------------

\section{Parameter definitions in the original coordinates}
\label{app:param-def-appendix}

This appendix provides explicit formulae for the parameters $b_l$, $c_l$, and $d_l$ that appear in the quartic polynomial \eqref{eq:drho-sqrd-3} and related expressions throughout Section~\ref{sec:mode-power}.
These parameters are expressed in terms of the system parameters $a_j$, $a_{j,k}$ and the conserved constants $\gamma_j$.

\begin{align}
N &= 2,\\
{b}_{0} &= {a}_{0} + \sum_{j=1}^{N} {a}_{j} {\gamma}_{j} + \frac{1}{2}\sum\limits_{j,k=1}^{N} {a}_{j,k} {\gamma}_{j} {\gamma}_{k}, \notag \\
{b}_{1} &= - \sum_{j=1}^{N} {a}_{j} - \frac{1}{2}\sum\limits_{j,k=1}^{N} \left({\gamma}_{j} + {\gamma}_{k}\right) {a}_{j,k}, \notag \\
{b}_{2} &= \frac{1}{N}\sum\limits_{j,k=1}^{N} {a}_{j,k}, \notag \\
{c}_{0} &= \prod_{j=1}^{N} {\gamma}_{j}, \notag \\
{d}_{0} &= {b}_{0}^{2} - 4 {c}_{0}, \notag \\
{d}_{1} &= 2 {b}_{0} {b}_{1}, \notag \\
{d}_{2} &= 2 {b}_{0} {b}_{2} + {b}_{1}^{2} - 4, \notag \\
{d}_{3} &= 2 {b}_{1} {b}_{2}, \notag \\
{d}_{4} &= {b}_{2}^{2}
\end{align}

% ---------- APPENDIX ------------
% --------------------------------

\section{Initial condition relations}
\label{app:init-conds}

This appendix gives the relationships between initial conditions for the physical system and the constants $\gamma_j$ and $\rho(0)$ used throughout the analysis.
These relations enable determination of the constants from specified initial values of $u_j(0)$ and $v_j(0)$.

\begin{align}
N &= 2, \notag \\
\rho{\left(0 \right)} &= - \frac{1}{N}\sum_{j=1}^{N} u_j{\left(0 \right)} v_j{\left(0 \right)}, \notag \\
\left. \frac{d}{d z} \rho{\left(z \right)} \right|_{\substack{ z=0 }} &= \prod_{j=1}^{N} u_j(0) - \prod_{j=1}^{N} v_j(0), \notag \\
{\gamma}_{j} &= u_j{\left(0\right)} v_j{\left(0 \right)} + \rho{\left(0 \right)}
\end{align}

% ---------- APPENDIX ------------
% --------------------------------

\printbibliography

\end{document}